\documentclass[letterpaper]{article}

\usepackage[T1]{fontenc}

\usepackage{geometry}
\usepackage{setspace}

\usepackage[style = chem-acs, doi=true, articletitle=true, url=true]{biblatex}
\usepackage{graphicx}
\usepackage{float}
\newfloat{scheme}{htbp}{los}
\floatname{scheme}{Scheme}
\floatname{chart}{Chart}
\newfloat{graph}{htbp}{loh}

\usepackage{chemformula} 
\usepackage[version = 4]{mhchem} 

\let\oldcite\cite
\newcommand{\normcite}[1]{\oldcite{#1}}
\usepackage{overpic}
\graphicspath{{figures/}}
\usepackage[final]{hyperref}
\hypersetup{
	colorlinks=true,       
	linkcolor=blue,        
	citecolor=blue,        
	filecolor=magenta,     
	urlcolor=blue         
}
\usepackage{subcaption}
\newcommand{\degree}{^\circ}
\usepackage{amsmath, amssymb}
\usepackage{xurl} 
\DeclareUnicodeCharacter{2212}{-}
\usepackage{wrapfig}
\usepackage{placeins}

\title{Active Learning for Tractable and Reproducible Pulsed Laser Deposition}

\usepackage{authblk}
\author[1,3]{Jackson S. Bentley}
\author[2]{Christopher Rouleau}
\author[2]{Ilia N. Ivanov}
\author[2]{T. Zac Ward}
\author[3]{Jiaqiang Yan}
\author[2,4]{Anghea Dolisca}
\author[3]{Rob G. Moore}
\author[3]{Gyula Eres}
\author[1]{Richard F. Haglund}
\author[2,*]{Sumner B. Harris}
\author[3,*]{Matthew Brahlek}

\affil[1]{Department of Physics and Astronomy, Vanderbilt University, Nashville, TN, 37235, USA}
\affil[2]{Center for Nanophase Materials Sciences, Oak Ridge National Laboratory, Oak Ridge, TN, 37831, USA}
\affil[3]{Materials Science and Technology Division, Oak Ridge National Laboratory, Oak Ridge, TN, 37831, USA}
\affil[4]{Biomedical Engineering Department, Carnegie Mellon University, Pittsburgh, PA 15213, USA}
\affil[*]{Corresponding author email: harrissb@ornl.gov and brahlekm@ornl.gov}
\date{}

\begin{document}

\maketitle


    

\begin{abstract}

This paper shows how data-driven machine learning approaches can improve growth control, reproducibility, and physical insight in the pulsed laser deposition (PLD) growth of correlated oxides. Despite well-known relationships between growth conditions and material properties, consistently producing high-quality films of complex materials like LaVO$_3$ remains difficult due to the highly non-equilibrium nature of PLD and the defects and competing phases that accumulate during growth. Here, we use an active learning framework based on Gaussian process Bayesian optimization that incorporates measured bulk and surface lattice properties along with impurity phase information to efficiently map the multidimensional growth space of LaVO$_3$ by PLD. By tuning the relative weighting of these properties, the model identifies an optimized region where phase-pure films of LaVO$_3$ exhibit two-dimensional surfaces, near-ideal lattice parameters, and minimal sub-band gap optical absorption. The trained model reveals clear competition among different defect formation mechanisms that are connected to unseen parameters like supersaturation and surface mobility, thus giving insight into the highly non-equilibrium process of PLD growth. Together, this demonstrates that property-guided machine learning can accelerate materials optimization while providing a new way to address fundamental growth mechanisms in PLD that enable understanding and utilization of quantum phenomena found in complex oxides.

\end{abstract}

\let\thefootnote\relax\footnote{Notice: This manuscript has been authored by UT-Battelle, LLC, under Contract No. DE-AC05- 00OR22725 with the U.S. Department of Energy. The United States Government retains and the publisher, by accepting the article for publication, acknowledges that the United States Government retains a non-exclusive, paid-up, irrevocable, world-wide license to publish or reproduce the published form of this manuscript, or allow others to do so, for United States Government purposes. The Department of Energy will provide public access to these results of federally sponsored research in accordance with the DOE Public Access Plan ( http://energy.gov/downloads/doe-public-access-plan ).}
\pagebreak

\section{Introduction}

Transition metal oxides (TMOs) are a playground for investigating fundamental physics and materials science, and they exhibit a range of functionalities that are relevant to computing, quantum sensing, and energy conversion technologies. Many of the functional properties arise due to complex orbital, charge and spin interactions which impact novel phenomena such as superconductivity, colossal magnetoresistance and metal-insulator transitions \cite{Dagotto2005}. Moreover, synthesizing materials as high-quality thin films is critical for harnessing these phenomena and facilitating new routes to tailor the novel electronic and magnetic states \cite{Miyasaka2006, Imada1998, Capone2002, Ohtomo2002, Okamoto2025}. The novel properties of TMOs have also translated into applications such as electrical switches, optical switches, active lasing media, and solar cells \cite{Hallman2021, Uchida2022, Wang2015}. Of interest here is the TMO perovskite LaVO$_3$ (LVO), which is Mott insulating with a Hubbard gap (1.1 eV) due to strong electron-electron correlations \cite{Arima1993VariationOxides}. LVO exhibits antiferromagnetic spin-ordering which coincides with antiferroic orbital-ordering, making it an interesting candidate for studying the competition between structural and spin coupling \cite{Miyasaka2003Spin-orbitalY}. The symmetry and ordering of the antiferromagnetic state also make LVO an interesting candidate altermagnet, which have been predicted to exhibit a spin splitting in the bandstructure and may be exciting systems for spintronics devices  \cite{Naka2025AltermagneticPerovskites}. Moreover, the strong electron-electron correlations make LVO a candidate for applications in heterointerface engineering, oxide electronics and solar cells \cite{Hotta2007, Assmann2013, Wang2015}. In fact, theoretical studies suggest that Mott insulators such as LVO can exceed the Shockley-Queisser limit on solar cell efficiency due to electronic correlations \cite{Manousakis2010, Petocchi2019}. Specifically, these studies predict that electron-electron scattering rates are an order of magnitude greater than electron-phonon scattering for moderate energy regions of the solar spectrum, amounting to lower thermal energy loss than conventional insulators \cite{Coulter2014}.

Although ternary TMOs, such as ABO$_3$ perovskites, offer a broad range of optical, electronic, and magnetic functionalities, synthesizing high-quality thin films in certain systems, such as LVO, remains difficult. These challenges stem from the need to control both cation stoichiometry and valence states, as well as understand the defects that form and their impact on properties. In MBE growth, cation ratios typically require precise manual control, though notable exceptions exist when one cation or a surface oxide is volatile, enabling adsorption-controlled growth \cite{10.1063/1.5023477, THEIS1998107, Brahlek2022WhatOxides}. Hybrid MBE (hBME) extends this concept by using metal-organics to create effective adsorption-controlled growth windows \cite{10.1116/1.3106610, Brahlek2022WhatOxides}. Valence control in MBE is generally achieved by adjusting the oxygen partial pressure, although pressure constraints for certain material systems can limit access to high valence states \cite{OH2005301, PhysRevB.71.052504, PhysRevMaterials.3.093401}. Alternatively, pulsed laser deposition (PLD) and sputtering provide simpler and more flexible growth approaches, often allowing nominal stoichiometry transfer of materials from the target to the growing film, but the process is highly non-equilibrium and complicates control over stoichiometry, valence states, and overall film quality.

LVO thin films have successfully been grown by MBE \cite{bairen2023exploring, PhysRevMaterials.4.064417}, hMBE \cite{Zhang2017, Brahlek2016}, sputtering \cite{CELINDANO201916658}, and PLD \cite{Hotta2006, Hotta2007, Wang2015, Wadehra2020, Zhang2021, Rotella2012, Rotella2015, Cheikh2024, Meley2018}. A variety of synthesis parameters by PLD have been reported for LVO [\autoref{table:PLD_param_table}]. Temperatures from 600 - 900 $\degree$C, oxygen partial pressures (PO$_2$) from 10$^{-7}$ - 10$^{-6}$ Torr, and fluences from 1.0 - 2.5 J/cm$^2$ have been reported to produce films with optimal structural, electronic, and optical properties. However, defects and thus growth conditions strongly affect these properties. For example, similar growth conditions can generate a wide range of out-of-plane lattice parameter ($c$) values from 3.935-3.985 \AA\ (for comparison this is limited to growth on SrTiO$_3$, where the optimum $c \approx$ 3.945-3.950 \AA). This assortment of $c$ values indicates the formation of point defects that originate from cation-cation and cation-anion non-stoichiometries \cite{Zhang2021, Zhang2017, Brahlek2016}, as well as the presence of impurity phases \cite{Hotta2006}. Moreover, stabilizing V in the desired V$^{3+}$ state is challenging since both the higher valence states of V$^{4+}$ and V$^{5+}$ are preferred and growth techniques like PLD typically utilize high-pressure ($\sim$1-100 mTorr) background gas to control the kinetics of growth species. The defects and impurity phases imparted during growth also have clear impacts on functional properties, such as optical absorption which can increase by a factor of two near the band edge \cite{Cheikh2024, Razavi2010, Rotella2015, Zhang2017}. Altogether, the variety of PLD growth parameters reported to produce optimal films comparable to those grown by hMBE (highly 2D surface, no impurity phases, $c$ $\approx$ 3.95 \AA\ and minimal sub-band gap absorption) conflicts with the reports demonstrating the extreme sensitivity of the structural and functional properties of LVO to growth conditions.

Autonomous experimentation through active learning has emerged as a powerful approach for accelerating materials discovery, optimizing synthesis procedures, and uncovering structure–property relationships \cite{Liu2022, Kusne2020, Harris2024, Mekki-Berrada2021}. In the area of materials physics, active learning methods typically integrate surrogate models, such as Gaussian processes (GPs), with Bayesian optimization. Gaussian process Bayesian optimization (GPBO) methods utilize initial samples and prior information to guide future sampling. Each new sample updates the model, reduces model uncertainty, and refines subsequent sampling decisions by balancing exploration of uncertain regions with exploitation of already identified local optima \cite{Frazier2015}. GPBO frameworks can drive both fully automated platforms as well as human-in-the-loop implementations in which autonomous data analysis and model-driven predictions augment experimental intuition. However, designing an experimental program that insightfully and tractably characterizes a growth parameter space and integrates active learning remains an active area of research.


Integrating active learning methods with PLD growth represents a promising direction due to the inherent complexity of the PLD process and the observation that nominally identical growth conditions often produce thin films with disparate properties. This lack of reproducibility is attributed to hidden or uncontrolled PLD growth variables. The most important fundamental PLD parameters are the supersaturation and the incident kinetic energy of the growth species \cite{Vasco_2006}. The importance of these parameters has been confirmed from single-shot growth data measured in real time by surface X-ray diffraction \cite{PhysRevLett.96.226104}. The analysis of these data has revealed a fast non-thermal component referred to as transient mobility characteristic of energetic PLD species and a slow thermal component of interlayer transport \cite{PhysRevB.84.195467}. A kinetic Monte Carlo model has independently reproduced the salient features of these single-shot PLD growth data \cite{PhysRevE.102.063305}. Single-shot real-time data related to the evolution of surface coverage during PLD growth has significantly advanced our understanding of the non-equilibrium growth mechanisms in PLD. However, direct correlation between experimental growth variables such as laser fluence, growth temperature and background pressure and their effect on the fundamental parameters has not been established. Ideally this should be done by performing real-time measurements, which are extremely demanding and require unique experimental facilities. Thus, the emergence of active learning methods offers a promising new approach for overcoming the lack of fundamental data by instead facilitating property-driven optimization of PLD growth, which may produce films with more desirable properties and give greater insight into the PLD growth process.

\begin{figure}[h]
  \centering
  \includegraphics[width=\linewidth,
                   trim=0cm 16.7cm 0cm 0.0cm, clip]{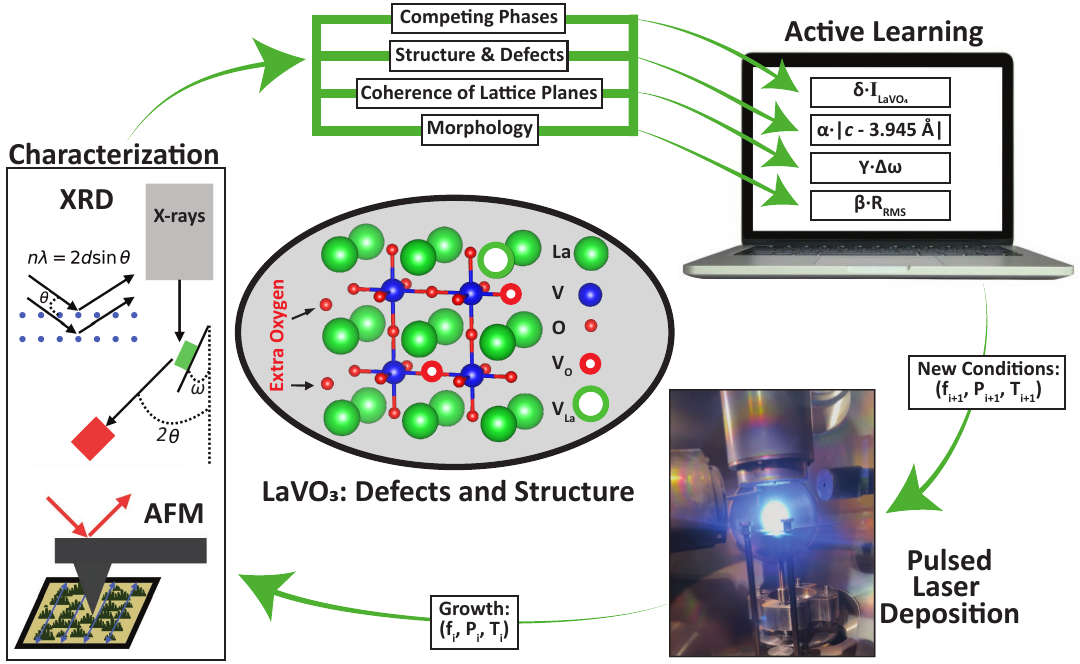}
  \caption{Illustration of the experiment workflow: The center ellipse illustrates the structure of LVO and common defects, as labeled. The rest of the figure depicts the cyclic workflow of this study. The workflow starts with synthesis by PLD, which is followed by characterization. Then, the results are fed into an active learning model, which suggests growth parameters for the next iteration.}
  \label{fig:Figure_1}
\end{figure}

Here, we demonstrate how data-driven machine learning methods can enhance growth control, reproducibility, and physical understanding in the PLD of the correlated TMO, LVO. To address the challenges of reproducibility in PLD and interpretable experimental design that integrates active learning, we utilize a human-in-the-loop framework, as illustrated in \autoref{fig:Figure_1}. This approach cyclically integrates machine learning, property characterization, and PLD growth to map a multi-dimensional parameter space, which enables efficient identification of optimal LVO growth conditions as well as deeper insight into the PLD growth process. A GPBO active learning model informed by bulk and surface lattice properties as well as impurity phases enables the interpolation of the growth quality landscape as a function of PO$_2$, substrate temperature, and laser fluence. This analysis reveals distinct regions that favor impurity LaVO$_4$ formation occur at high pressures and high temperature, and regions that favor point defect formation occur at low pressures and low temperature. Varying the relative contributions of measured properties on the GPBO model highlights an optimization valley where phase-pure, defect-minimized films with smooth 2D surfaces and near-ideal lattice parameters form. These films show minimal sub-band gap optical absorption, similar to MBE-grown films. Within our experimental bounds, we consistently achieve these optima, which demonstrates that property-guided machine learning integrated with complex growth techniques can accelerate and deepen understanding of phase formation and defect control in correlated oxide systems.

\subsection{Workflow and Objective Function}

As highlighted in \autoref{fig:Figure_1}, the workflow started with sample growth using conditions common for PLD and reported in the literature. For each film, measured values of the $c$, root-mean-square roughness ($R_{RMS}$), full-width half-maximum of the rocking curve of the (002) film peak ($\Delta\omega$), and integrated intensity of the impurity LaVO$_4$ phase ($I_{LaVO_4}$) were used to quantify bulk and surface film properties. As indicators of film quality, these quantities provided critical information on defect concentration and stoichiometry, surface morphology, crystalline coherence, and competing phases, respectively. Although each type of measurement quantifies film properties, it is not immediately clear how they collectively describe a global synthesis parameter space spanned by PO$_2$, temperature and laser fluence. Nor is it clear how they provide insight into the processes that dictate the topography of the parameter space and positions of optimum regions. The GPBO framework outlined here begins to address these challenges. The framework translates measured quantities into an objective function, which characterizes the synthesis parameter space and can be the basis for gaining deeper understanding of the PLD growth process. The objective function is the sum of the deviation from the target $c$ (3.945 \AA), $R_{RMS}$, $\Delta\omega$, and $I_{LaVO_4}$, which are weighted by the constants $\alpha$, $\beta$, $\gamma$, and $\delta$, respectively:
\begin{equation}
    y_i(f, T, PO_2) = \alpha \cdot |c_{i} - 3.945| + \beta \cdot R_{RMS, \ i} + \gamma \cdot \Delta\omega_i + \delta \cdot I_{LaVO_{4}, \ i}.
    \label{equation_obj}
\end{equation}

From the objective function (\autoref{equation_obj}), each sample acquires a score $y_i$, which serves as input for the calculation of the GP surrogate model mean, variance, and acquisition function. The calculation of the GP surrogate model mean interpolates the value of the objective function throughout the growth parameter space, thereby quantifying the film quality landscape. The calculation of the variance gives the uncertainty of the GP mean. The acquisition function quantifies the expected improvement to the GP mean due to another growth iteration. That is, the acquisition function informs the selection of growth parameters for the next experiment. (See \autoref{math_GPBO} for a mathematical description of the GP mean, variance, and acquisition function.) Lower overall scores ($y_i$) and individual measurement values (|$c - 3.945$|, $R_{RMS}$, $\Delta\omega$, and $I_{LaVO_4}$) correspond to higher quality samples, where 3.945 \AA \ is the target $c$. (More discussion of the target $c$ is included later and in \autoref{poisson_c}.) To include $c$ in the objective function, this term is the absolute value of the difference of the measured $c$ and the target $c$, which maintains the correspondence of lower scores and higher quality samples. 

The choice of the parameter weights (\textit{i.e.} the values of $\alpha$, $\beta$, $\gamma$, and $\delta$) reflected the magnitude and distribution of values for the particular parameter and their relative importance. Explicitly, the difference between the minimum and maximum measured values (|$c - 3.945$|, $R_{RMS}$, $\Delta\omega$, and $I_{LaVO_4}$), so called min-max scaling, was used to determine the magnitude of each weight, $\alpha$, $\beta$, $\gamma$, and $\delta$ in \autoref{equation_obj}. As an example using the $\Delta\omega$ term, setting $\gamma$ = $\frac{1}{\Delta\omega_{max} - \Delta\omega_{min}}$ makes the term $\gamma\cdot\Delta\omega_i$ unitless and also of order unity. Although this construction applies to the calculation of the other weights, $c$ encodes the most information about overall crystal structure and quality. Also, measurements of $I_{LaVO_{4}}$, $\Delta\omega$, and $R_{RMS}$ were moderately correlated, while deviation from the target $c$ was not correlated to the other individual metrics, as shown in \autoref{fig:supp_corre} and discussed later in the paper. Therefore, the value obtained from applying min-max scaling to measurements of |$c - 3.945$| was multiplied by two to obtain the weight $\alpha$. This construction balances the objective function and accounts for the importance of $c$  as a measure of structure and local chemical environment.

\section{Results and Discussion}

\subsection{Individual Characterization Measurements}

\begin{figure}[h]
  \centering
  \includegraphics[width=\linewidth,
                   trim=0cm 17.4cm 0cm 0cm, clip]{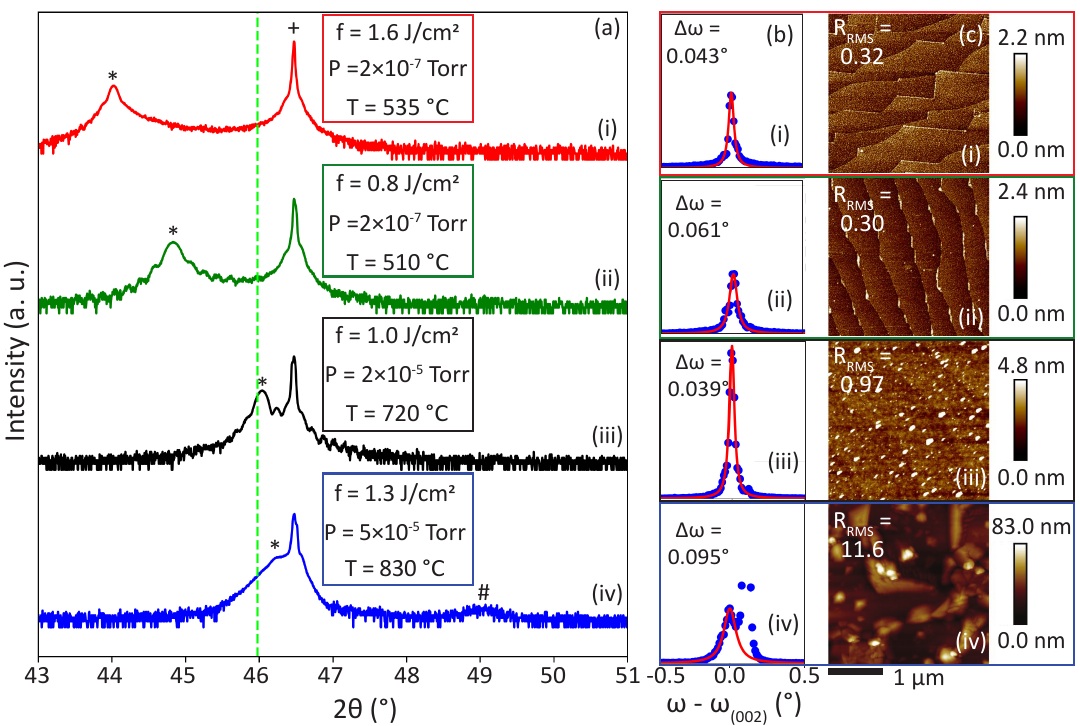}
  \caption{(a) Sample XRD 2$\theta$-$\omega$ scans for several samples at different growth conditions. The dotted green vertical line indicates the XRD 2$\theta$ peak position that corresponds to the target $c$, the * symbol indicates the LVO (002) peak, the \# symbol indicates the LaVO$_4$ (040) XRD peak, and the + symbol indicates the STO (002) peak. The labels (i-iv) indicate specific samples. (b) Rocking curve measurements of the LVO (002) peak of each sample. The solid red line is the best fit using a pseudo-Voigt function to quantify $\Delta \omega$. (c) AFM images of the surface morphology. The $R_{RMS}$ values are in units of nm.}
  \label{fig:Figure_2}
\end{figure}

Results in \autoref{fig:Figure_2} summarize the characterization methods used to extract |$c - 3.945$|, $I_{LaVO_4}$, $\Delta\omega$, and $R_{RMS}$ from films grown under varying PO$_2$, laser fluence, and temperature. \autoref{fig:Figure_2} (a) shows X-ray diffraction (XRD) 2$\theta$-$\omega$ scans for this set of samples. The sharp peak around 46.5$\degree$ is the STO (002) peak (marked with a + symbol), and to the left (lower 2$\theta$) is the LVO (002) peak (indicated by an $*$ symbol). The vertical green line indicates the 2$\theta$ position for ideal LVO strained to STO: $c\approx$ 3.945 Å (2$\theta \approx$ 46$\degree$). These data show that the LVO (002) peak moves dramatically by up to 2$\degree$ based on growth conditions, indicating the high sensitivity of $c$ to the accumulation of defects. For all of the films grown here, $c$ either increased or slightly decreased at non-ideal growth conditions. For example, at low temperature and low pressure [top red curve, sample (i) in \autoref{fig:Figure_2} (a)], the LVO peak shifts to lower 2$\theta$, and the oscillations are absent, indicating poor quality and the accumulation of a large amount of defects. Conversely, for higher pressure and temperature [sample (iv) in \autoref{fig:Figure_2} (a)], the LVO peak shifts slightly to larger 2$\theta$, the peak-shape becomes distorted, and the oscillations vanish, which also indicates poor sample quality. In between these conditions, the 2$\theta$ position for the peak ranges from $\sim$44.5-46.0$\degree$ [sample (ii) and (iii)]. Their peak shapes are symmetric and exhibit clear thickness oscillations due to the coherent scattering from the top and bottom film surface, which indicate the films are near-atomically flat. Without \textit{a priori} knowledge of ideal lattice parameters, this indicates that $c$ and peak shape alone are insufficient to identify optimum conditions.

The out-of-plane lattice parameter encodes information about point defects and impurity phases, which tend to increase $c$ for most perovskites \cite{https://doi.org/10.1002/adfm.201702772}, but $c$ has been reported to either increase or decrease for non-stoichiometric LVO grown by hMBE and PLD \cite{Zhang2017, Zhang2021}. Since LVO is under compressive strain, a reduction of $c$ can also indicate that the film has epitaxially relaxed relative to STO, which convolutes the effect of defects and how they are accommodated by the lattice. The distortion of the lattice is unique to each specific defect, and, thus, likely to be unique to the growth temperature and method. As such, the choice of 3.945 Å as the target $c$ followed from literature values of PLD growth \cite{Rotella2012, Vrejoiu2016}. And it followed from an estimate of $c$ due to compressive strain using a bulk lattice parameter of $\sim$3.925 \AA \ and Poisson ratio values from the literature (0.3 - 0.4) \cite{Brahlek2016, 10.1063/1.5045704} [see \autoref{poisson_c} and \autoref{fig:supp_poisson}]. For PLD growth of LVO, generally lower fluence and higher PO$_2$ lead to near ideal measured values of $c$ (typically lower $c$ values), reflecting the effect of closer-to-ideal cation-cation stoichiometry. This roughly seems to be borne out in the data in \autoref{fig:Figure_2} (a), by comparing the green curve (lower temperature and pressure) and the black curves (higher temperature and pressure). Moreover, as shown in \autoref{fig:htxrd_before_after_no_Pt} and \autoref{fig:htxrd_many_scans} for high-temperature annealed LVO, the effect of oxygen vacancies on $c$ can be eliminated since vacancies would require $V\xrightarrow{} V^{2+}$, which would have high formation energies.

The highly distorted peak shape of the sample (iv) shown in blue in \autoref{fig:Figure_2} (a), correlates with the appearance of a secondary peak at around 2$\theta\approx$ 49$\degree$. This peak is consistent with LaVO$_4$ formation, as marked by a \# symbol in \autoref{fig:Figure_2} (a). The appearance of crystalline LaVO$_4$ in this sample indicates that these growth conditions are near the phase boundary for the formation of V$^{5+}$. In fact, V$^{5+}$ spontaneously forms on the surface of LVO films postgrowth due to oxidation in the air \cite{10.1063/1.2775889}. Similarly, during PLD growth LaVO$_4$ formation should occur at higher PO$_2$ and low temperature which reflects the thermodynamic expectation for oxidation. Hotta et al. found this rough trend where LaVO$_4$ forms at PO$_2$ higher than 10$^{-6}$ Torr, yet found little effect due to temperature \cite{Hotta2006}. They conjecture that interfacial energies dominate over thermodynamic considerations for the bulk, which allows for the formation of the perovskite phase even at high temperature. As discussed later, the approach here shows the phase diagram is far more rich.

The width of the rocking curve taken by rotating the sample about the vector normal to the scattering plane ($\theta$) quantifies the coherency of the (00L) lattice planes, and is therefore a more sensitive metric to the long range disorder such as impurity phases. Rocking curve widths for this sample set are shown in \autoref{fig:Figure_2} (b), where the widths range from a maximum of $\Delta\omega\approx$ 0.095$\degree$ to a minimum of 0.039$\degree$. The lowest $\Delta\omega$ occurs for the film with $c$ closest to ideal. And interestingly the sample indicated by the red curve, which had nominally worse $c$ and peak character, had a better $\Delta\omega$ compared to the sample indicated by the green curve. Finally, the bottom curve, clearly had the largest rocking curve width (it is noted that the $\Delta\omega$ scan had a double peak due to the proximity to the STO (002) peak position). 

Film roughness acquired through atomic force microscopy (AFM) measurements captures surface uniformity and reflects the size distribution of crystallographic domains. \autoref{fig:Figure_2} (c) shows representative AFM measurements from the same four samples shown in \autoref{fig:Figure_2} (a) and (b). The large $R_{RMS}$ of sample (iv) indicates poor coupling between the film and substrate, \textit{i.e.} non-epitaxial growth, which drives the formation of growth fronts unrelated to the substrate orientation and structure. In fact, LaVO$_4$ formation was prominent in this sample [as illustrated in (a)]. Therefore, the large $\sim$50 nm tall features were likely LaVO$_4$ intergrowths. In the PLD of SrMoO$_3$, another TMO where it is difficult to stabilize a low valence state on the B site, lower temperatures generally lead to flatter surfaces \cite{Alff2014}, and these samples follow this trend. Not only does roughness encode information about growth fronts and domain sizes, but it is also important for heterostructure engineering, which highlights the importance of optimizing this parameter.

\subsection{Growth Iteration Process, Final GP Surrogate Model, and Modeling from Individual Characterization Measurements}

\begin{figure}[H]
  \centering
  \includegraphics[width=\linewidth,
                   trim=0cm 21.7cm 6.1cm 0cm, clip]{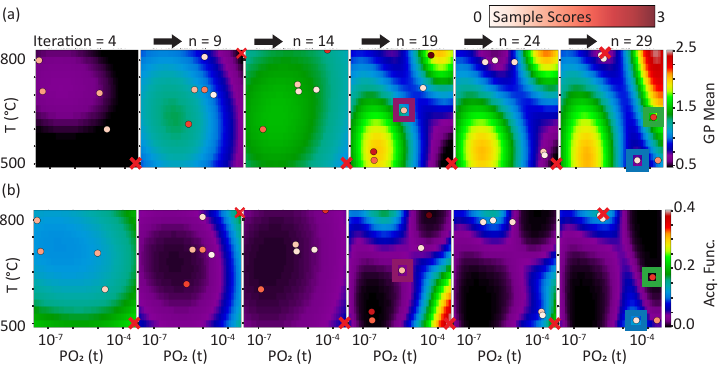}
  \caption{(a, b) Evolution of the mean (a) and acquisition function (b) of GP surrogate model showing the projection of the mean of the GP surrogate model onto the deposition temperature ($\degree$C) and PO$_2$ (Torr) plane with increasingly more samples, as indicated at the top. White-to-red circles indicate sample scores. The red Xs indicate the maxima of the acquisition function after the last growth iteration in each column. The red X for growth iteration 29 is in the same position as the global optimum. The unfilled blue, purple and green squares correspond to the growth conditions of the samples given by the blue, purple and green curves in \autoref{fig:Figure_5_labels}.}
  \label{fig:Figure_3}
\end{figure}

\autoref{fig:Figure_3} (a) and (b) show the evolution of the GP mean of the surrogate model (a) and the acquisition function (b) as additional samples were grown and measured. These plots are a 2D projection of the 3D growth parameter space (temperature, PO$_2$, and fluence) onto the temperature-PO$_2$ plane, which were found to be the more sensitive of the three variables. The GP mean of the surrogate model [\autoref{fig:Figure_3} (a)] illustrates the film quality landscape as a function of temperature and PO$_2$, where cool colors (lower values) indicate good sample quality and warm colors (higher values) indicate poor sample quality. The acquisition function [\autoref{fig:Figure_3} (b)] quantifies the expected improvement of the GP mean for another sample growth. In other words, the warm-to-cool colors indicate large-to-small expected improvement to sample quality for the next growth iteration. In \autoref{fig:Figure_3} (a) and (b) going from left to right, additional samples were added, where the first column was created using 4 samples and the last one was the cumulative result of all 29 samples. The samples added during each iteration are indicated by the circles, where the color indicates the sample score and lower scores equal better quality, as in \autoref{equation_obj}.

After the first few growth iterations (4, 9, and 14), the GP mean shows a wide maximum and does not change much with the additional samples. However, after 19 growth iterations, particularly after a sample growth at high PO$_2$ and temperature, the film quality landscape changed significantly, as additional samples were grown in this unexplored and non-optimum part of the phase space. Specifically, the change to the landscape entailed the wide maximum shifting toward lower pressure and temperature and a second maximum appearing at high pressure and temperature. Surprisingly, these maxima are separated by a wide valley and two local minima. The GPBO algorithm predicts optimal film growth at moderate PO$_2$ ($\sim$10$^{-6}$ - 10$^{-5}$) and high temperature after iteration 19, where the global optimum is eventually identified (at the red X for growth iteration 29). The acquisition function tends to mirror the GP mean by frequently predicting the most improvement to sample quality will occur near the global optimum of respective growth iterations. After 24 growth iterations, adding a couple growths at large PO$_2$ and low temperature as well as the eventual global optimum caused the GP mean to stabilize at the final mapping indicated by growth iteration 29.

The final GP surrogate model provides information on the growth conditions that drive film properties. The global optimum is at temperature = 820 $\degree$C, fluence = 0.8 J/cm$^2$, and PO$_2$ = 1.2$\times$10$^{-6}$ Torr; another near optimal growth regime is at temperature = 500 $\degree$C, fluence = 0.8 J/cm$^2$, and PO$_2$  = 2.4$\times$10$^{-5}$ Torr. Interestingly, a valley of growth conditions with low GP mean values links the two areas, all of which is illustrated in \autoref{fig:Figure_3} (a) iteration 29. Since higher temperatures along this valley require lower PO$_2$ for film growth with minimal defects, these results contravene expectations from the thermodynamics of oxidation reactions where higher temperatures are expected to result in oxygen loss. One resolution is that at high temperatures the STO may act as an oxygen source, which compensates for the reduced PO$_2$ \cite{Schneider2010, Spinelli2010}. Also, taking the gradient of the final GP surrogate model mean reveals that PO$_2$ drives the topography of the growth landscape more than temperature or fluence. The gradient is largest along the PO$_2$ direction, and this fact supports the conclusion that PO$_2$ plays a role in driving LaVO$_4$ formation and likely a role in driving defect formation [see \autoref{supp_gp_grad}]. In this way, the implementation of active learning clearly demonstrates subtleties of the PLD process that are crucial to optimal LVO growth.

\begin{figure}
  \centering
  \includegraphics[width=\linewidth,
                   trim=0cm 5cm 0cm 5cm, clip]{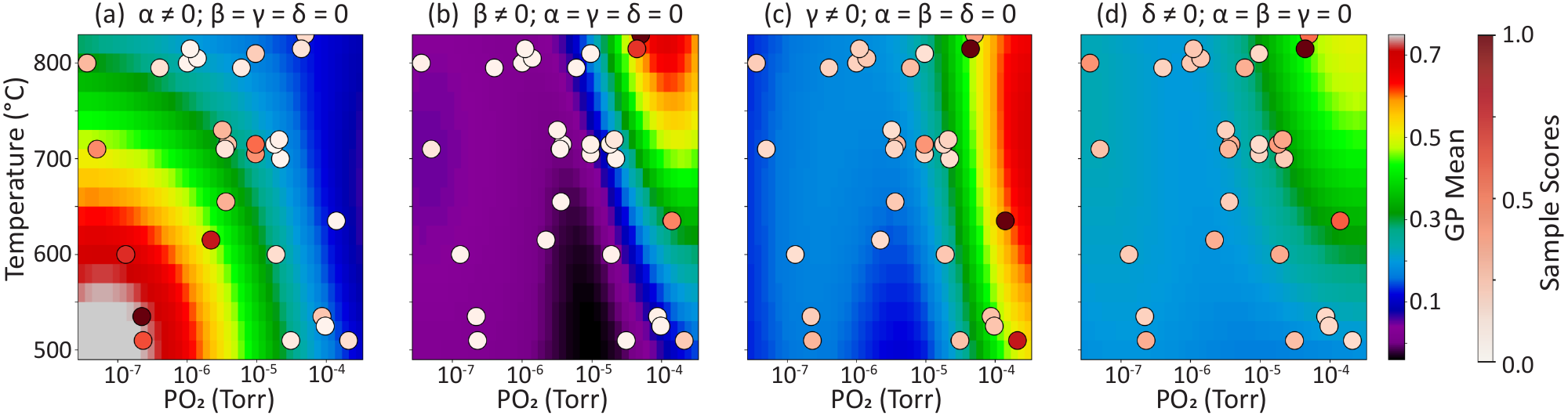}
  \caption{ (a-d) 2D projections onto the deposition temperature and PO$_2$ plane of the final GP surrogate from Fig. 3, but considering only a single type of measurement. Setting one of the weights ($\alpha$,  $\beta$, $\gamma$, or $\delta$) equal to the value obtained by min-max scaling and the rest equal to zero, as indicated at the top of each panel, creates the plots. Plots of $\alpha$,  $\beta$, $\gamma$, and $\delta$ $\neq 0$ quantify film structure and defects, morphology, coherence of lattice planes, and the prevalence of the LaVO$_4$ phase, respectively. Lower GP mean values correspond to higher quality growth regions, and lower sample scores correspond to higher quality samples.} 
  \label{fig:Figure_ind}
\end{figure}

Although the weighted sum of the individual measurements enabled sample optimization through the objective function (\autoref{equation_obj}), the individual measurements provide insight into the mechanisms that drive the LVO growth landscape pictured in \autoref{fig:Figure_3} (a) growth iteration 29. Recalculating the objective function to include only one type of measurement (|$c$ - 3.945|, $R_{RMS}$, $\Delta\omega$, or $I_{LaVO_4}$) can elucidate the defect each measurement probes and how those defects impact the global growth landscape found in \autoref{fig:Figure_3}. \autoref{fig:Figure_ind} shows the 2D projection onto the temperature-PO$_2$ plane with all 29 samples from the plots in \autoref{fig:Figure_3}. Here, the GP mean was calculated by setting all the weights to zero except one, as indicated in the label at the top of each plot. The sample scores, indicated by white-to-red circles, are the result of min-max scaling measured values by the respective weight not equal to zero. For $\alpha \neq 0$ and $\beta = \gamma =\delta =0$ [\autoref{fig:Figure_ind} (a)], the GP mean shows a single large maximum at low pressures and low temperatures, similar to the local maximum in the same region in \autoref{fig:Figure_3} (a). Conversely, the corresponding plots with $\beta \neq 0$ (b), $\gamma \neq 0$ (c), and $\delta \neq 0$ (d) show qualitatively similar behavior with a single maximum at high pressure and high temperature, reflecting the second local maximum in \autoref{fig:Figure_3} (a). 

The differences among these plots highlight two separate defect formation mechanisms that manifest distinctly in the physical measurements of $c$, $R_{RMS}$, $\Delta\omega$, and $I_{LaVO_4}$. Since $\alpha$ is the weight that scales $c$, it is clear that the lower maximum is caused by the accumulation of structural disorder that alters $c$ alone, most likely point defects. Such defects are likely associated with damage induced by high-velocity plume species at low pressure, and the low thermal mobility is insufficient to rearrange the adatoms to heal defects. Thus, increasing temperature likely reduces the incorporation of these defects by the lattice. Also, increasing the pressure, reduces the kinetic energy of plume species by increasing the energy transfer from the ablated material to the gas, thus resulting in less kinetic defects. These two aspects together create the local maximum seen in \autoref{fig:Figure_ind} (a). Similarly, \autoref{fig:Figure_ind} (b-d) highlight the second local maximum, which clearly originates from LaVO$_4$ formation as directly measured by $I_{LaVO_4}$. In fact, the measurements of $I_{LaVO_{4}}$, $\Delta\omega$, and $R_{RMS}$ were found to be moderately correlated, while deviation from the target $c$ was not correlated to the other individual metrics, as shown in the \autoref{fig:supp_corre}. Therefore, impurity phases have a dramatic effect on measurements of $R_{RMS}$ and $\Delta\omega$ since inclusions of a secondary phase have a strong impact on lattice coherence and smoothness of the surface, whereas these measurements have little sensitivity to point defects. Also, LaVO$_4$ formation is favored at higher oxygen pressures, but counterintuitively is favored at higher temperatures as well. Currently, we do not have a simple understanding for this behavior, but it likely reveals complexities of how non-equilibrium processes such as the plume dynamics affect the oxidation process of vanadium on the growing surface \cite{Geohegan1996}.

Taken together, the compilation of the \autoref{fig:Figure_ind} (a-d) gives a clear rationale for the valley of near-optimum conditions with an optimum on both ends, as shown in \autoref{fig:Figure_3} (a). The optimum conditions are the result of the competition among point-defects that have a strong impact on $c$ and phase segregation caused by over oxidation. These results offer a comparison to the previous systematic study by Hotta et al. of LVO film growth by PLD as a function of temperature and PO$_2$, which found a far simpler behavior \cite{Hotta2006}. The previous result indicated that there are distinct pressure regimes where phase-pure LVO or LaVO$_4$ forms, and a wide intermediate regime where there is phase coexistence. They identified little temperature dependence. These findings were based on surface measurements, AFM and layer oscillations of reflection high-energy electron diffraction (RHEED), as well as XRD to identify phase purity. The optimal regions for the three plots with $\beta, \ \gamma, \ \delta \neq 0$ in \autoref{fig:Figure_ind} align roughly with their results, indicating their optimization was approximately based on phase purity. 

The GPBO approach shown here gives a much clearer and objective view of the larger data set and thus deeper insight into the physics at play while also addressing the complexities inherent to PLD growth. Interestingly, their optimum conditions were at $\sim$600 $\degree$C and 10$^{-6}$ Torr, and the onset of a competing LVO-LaVO$_4$ growth regime was at an order of magnitude lower PO$_2$ than this study. This significant difference likely stems from several factors. First, they used a LaVO$_4$ instead of a LaVO$_3$ target, which likely sourced more oxygen (either free oxygen or more oxygen prebonded to the cations in the plume), thereby increasing the effective PO$_2$. Secondly, this discrepancy could be a difference in chamber configuration and differences in measurements of temperatures and pressures, which, although very precise, may be inaccurate measures of actual sample temperatures and pressures at the sample or target locations. Also, since STO desorbs some oxygen at high temperature and can act as an oxygen source, differences in substrate preparation, particularly annealing temperatures and duration, likely shift the GP surrogate model. The GPBO approach helps address these discrepancies by objectively mapping out the growth landscape, thereby revealing differences that can originate from factors such as target character, details of the substrate preparation, or unique instrumental configurations. Such inconsistencies further highlight the value of high-throughput experimental phase mappings all taken from a single growth chamber, as they generate consistent, high-quality ground-truth datasets suitable for machine learning. In highly non-equilibrium techniques such as PLD, this approach using property-driven optimization is particularly valuable since nonlinear interactions and hidden or uncontrolled growth variables can drive large variations in film properties.

\section{Sample at Global Optimum and Applications}

\begin{figure}[h]
  \centering
  \includegraphics[width=\linewidth,
                   trim=0cm 12.2cm 0cm 1.5cm, clip]{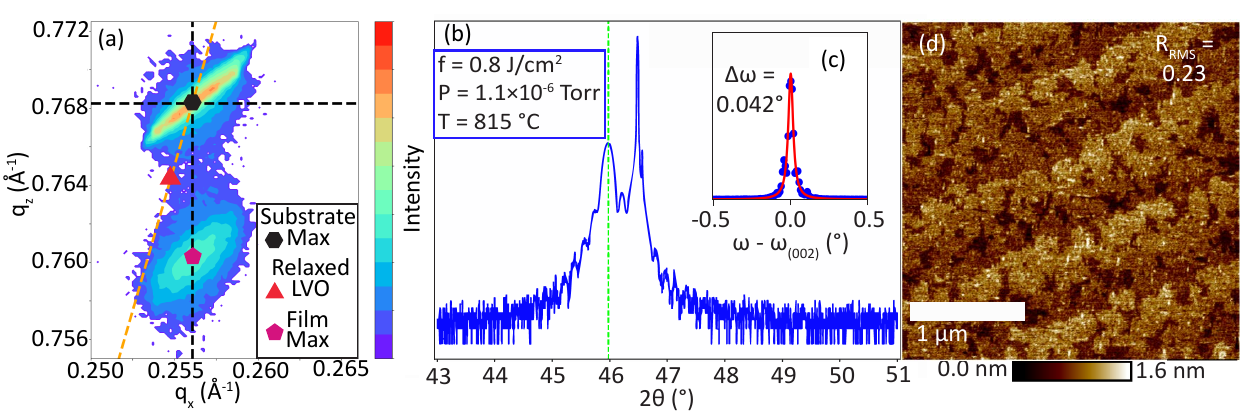}
  \caption{Characteristics of film grown at global optimum: (a) Reciprocal space map about the (103) reflection in STO. The red triangle denotes the position of unstrained LVO. The dotted black line specifies the substrate peak position, and the dotted orange line indicates peak positions of unstrained or bulk lattice parameters. (b) XRD 2$\theta$-$\omega$ data. The dotted, vertical line indicates the 2$\theta$ position that corresponds to the target $c$. (c) Rocking curve measurements of the LVO (002) film peak on the same scale as \autoref{fig:Figure_2} (b). (d) AFM measurement for the sample with $R_{RMS}$ in nm.}
  \label{fig:Figure_4}
\end{figure}

For growth iteration 29, we grew a sample near the predicted global optimum: temperature = 820 $\degree$C, fluence = 0.8 J/cm$^2$, and PO$_2$ = 1.2$\times$10$^{-6}$ Torr. In \autoref{fig:Figure_4} (a), a reciprocal space map about the STO (103) reflection shows the optimal film is strained to the substrate since the substrate and film maxima have the same q$_x$ value. Visualized with the dotted black lines, this relationship indicates pseudomorphic growth and an epitaxial relationship to the substrate. The red triangle in \autoref{fig:Figure_4} (a) indicates the peak parameters of bulk or completely relaxed LVO, assuming a cubic unit cell with $c$ = a$_{\parallel}$ = 3.925 Å. The dotted orange line is a guide to the eye and connects the STO maximum to the peak of bulk LVO in reciprocal space. As shown in \autoref{fig:Figure_4}, this sample has an $c$  = 3.947 Å (b), $I_{LaVO_4}$ = 6.2 (\textit{i.e.} no resolvable signature of LaVO$_4$) (b), $\Delta\omega$ = 0.042$\degree$ (c), and $R_{RMS}$ = 0.23 nm (d). These data together show that this sample grown at the predicted optimum conditions has high structural quality, consistent with bulk crystals and the highest quality hMBE grown films. 

\begin{figure}[H]
\centering 
\begin{minipage}{0.67\linewidth}
    \centering
    \includegraphics[width=\linewidth, trim=0cm 8.5cm 0cm 0.8cm, clip]{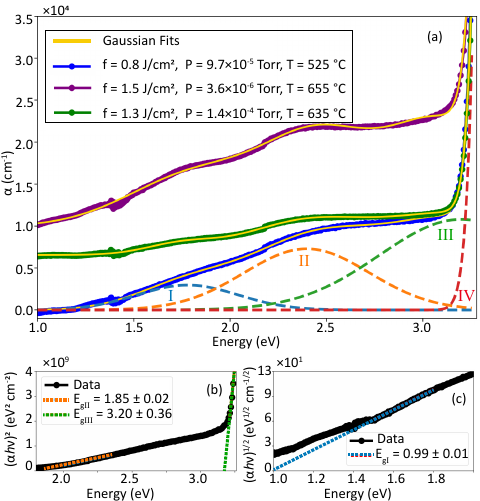} 
\end{minipage}
\hfill
\begin{minipage}{0.32\linewidth}
    \caption{(a) Absorption spectra for three different samples are the thick purple, green, and blue lines. The growth conditions for the samples with an absorption spectrum are listed in the legend and boxed in \autoref{fig:Figure_3} (a, b). Thin dashed lines are individual Gaussian fits to each peak from the absorption spectrum given by the blue curve, and the combined fits are the thin solid yellow curves. The blue (I) and orange (II) dotted lines correspond to V \textit{t}$_{2g}$ → V \textit{t}$_{2g}$ transitions, and the green (III) and red (IV) dotted lines correspond to O 2p → V \textit{t}$_{2g}$ transitions. (b-c) Tauc plots for direct (b) and indirect (c) interband transitions that correspond to absorption data from the blue curve in (a), the sample from the low-temperature growth mode. The legends contain the electronic transition gap (b) and band gap (c) values obtained from linear fits.} 
    \label{fig:Figure_5_labels}
\end{minipage}
\end{figure}

To further explore sample properties, absorption spectra illustrate electronic transitions and contain information about defect concentration and phase purity \cite{Zhang2017, Zhang2021, LaVO4_gap}. The absorption coefficient $\alpha$ (cm$^{-1}$) gives the amount of light absorbed per unit length, which is not only sensitive to material quality but also establishes an upper limit on how much light energy can be converted to electronic work. Testing the effects of growth parameters on absorption, we grew samples near three different points that yielded various scores from individual measurements during the growth iteration process: the low-temperature growth mode (blue curve), a point with an $c$ of 3.992 Å and low measured values otherwise (purple curve), and a point with the same $c$ as the low-temperature growth mode but large measured values otherwise (green curve). The unfilled blue, purple, and green squares in \autoref{fig:Figure_3} indicate the growth parameters of the samples given by the blue, purple, and green curves, respectively. Shown in \autoref{fig:Figure_5_labels} (a), absorption spectra for all three films largely match the absorption peaks found in the literature, except for the peak near 3 eV (O 2p → V \textit{t}$_{2g}$ transition) \cite{Zhang2017, Wang2015}. Compared to the literature, these results show a larger spectral weight for this transition, where we would expect impact ionization which would explain its increased spectral weight and the uncertainty in the transition into the E$_{g2}$ states. But peak broadening could also account for the large spectral weight and gap uncertainty for the peak near 3 eV. The absorption data for the low-temperature growth mode (blue curve) shows little sub-band gap absorption, but broad peaks. Sub-band gap absorption and broad absorption peaks generally correspond to the presence of defects, resulting in ambiguous conclusions about electronic structure and defect concentration. Tauc plots of the film grown at the low-temperature growth mode, shown in \autoref{fig:Figure_5_labels} (b) and (c), match electronic transition and band gap values from \normcite{Wang2015}, except for the third gap (3.2 eV), though it is within the uncertainty. \autoref{fig:Figure_5_labels} (b) shows a Tauc plot for the direct transitions, and (c) shows the fit to the indirect band gap. In \autoref{tauc_b41} and \autoref{tauc_b42}, Tauc plots for the other two films demonstrate significant sub-band gap absorption with band gaps near 0 eV, but the two direct electronic transition gaps largely match values from the low-temperature growth mode and the literature. This together shows that even though optical absorption was not explicitly incorporated into the GP mean model, the optimum conditions produced samples with the cleanest spectra matching those of hMBE grown films \cite{Zhang2017}. Further work that incorporates optical or other functional measurements into the active learning model would likely reveal the dependence of fundamental electronic behavior on the accumulation of particular defects. 

\section{Conclusions and Future Studies}


In summary, we establish a property-guided active-learning framework for optimizing the PLD of a correlated oxide, LVO, that efficiently navigates a complex growth landscape while revealing the competing defect mechanisms governing film quality. The framework enables convergence to optimal growth conditions within the bounds of the experimental parameters. Physically, this strategy optimizes growth conditions by fingerprinting both the conditions where defects arise and the effects of specific defects on film properties. By mapping defects and phase impurities to growth conditions, this active-learning approach could likewise enable transient absorption studies that pinpoint the experimental signatures of defects and phase impurities, which alter thermalization. The present study also suggests a route to implementing LVO in oxide electronics, mapping a large parameter space to highlight a low-temperature, moderate-PO$_2$ growth mode. This may be the only optimal, accessible approach in cases where experimental constraints limit maximum substrate temperature or minimum PO$_2$. Lower-temperature regimes are more attractive for industrial applications, for example. 

Finally, this study demonstrates an approach to increase the reproducibility of nominal synthesis conditions for the PLD of LVO. The GP surrogate model in \autoref{fig:Figure_3} and \autoref{fig:Figure_ind} facilitates comparison to other studies by mapping film properties onto the growth parameter space. And since the calculation of the GP surrogate model assumes some noise generation at each measurement, it is possible to objectively identify outlier growth results by the distance from measured values to the GP surrogate model, thereby providing a way to quantify reproducibility of PLD-grown films. Property-driven optimization also addresses the intrinsic constraints of PLD growth because non-equilibrium interactions govern growth but are difficult to link to accessible experimental growth variables, such as temperature, pressure, and fluence. By focusing on measurement outcomes instead of incompletely understood growth variables, this data-driven approach mitigates apparent inconsistencies and sample-to-sample variability. Moreover, this study underscores the value of high-throughput property characterization of samples grown in the same chamber, yielding reproducible datasets suitable for use as ground truth in machine learning models.

In future studies of LVO, this procedure can guide further exploration of the deposition phase space to synthesize higher quality material. For example, introducing H$_2$ + Ar background gas can lead to much lower oxidation rates and therefore mitigate the creation of defects from energetic plume species while also stabilizing V$^{+3}$ and suppressing LaVO$_4$ formation. Introducing multiple rare-earth cations to the A site is another strategy for reducing the impact of defects on the local chemical environment and lattice structure \cite{Mazza2024EmbracingDesign, Brahlek2022WhatOxides}. As in other material systems, alloying AVO$_3$ manages defects while conserving desirable physical properties and representing a design parameter for orbital or magnetic ground states \cite{Yan2024OrbitalOxides}.

Incorporating complementary probes of structure and electronic properties --- such as transmission electron microscopy, transient absorption, ARPES, and X-ray spectroscopies of chemical states by RHEED --- could also elucidate the way that defects and impurity phases impact relaxation pathways and electronic structure. Especially exciting is the possibility of incorporating machine learning with \textit{in situ} probes such as RHEED, which offer high throughput measurements that can enable real-time optimization of film properties \cite{Lapano2019ScalingSilicon, https://doi.org/10.1002/adfm.201702772}. At any rate, the synthesis of higher quality LVO films can lead the way to more precise experiments on correlation physics, illustrate growth strategies for perovskites with low-valence B-sites, and illuminate design strategies for integrating LVO into oxide electronics.

Looking beyond its implementation in this work, the active-learning framework is in principle agnostic to material system as well as growth method and could integrate different surrogate models, acquisition functions, or objective functions. Optimizing synthesis here involved minimizing competing phases, point defects, mosaicity, and surface inhomogeneities, and high-quality synthesis for other material systems must address the same problems. Applying the same framework to a transition metal dichalcogenide, for example, might require another term to be included in the objective function, like the lineshape of Raman spectra since that measurement is sensitive to defects in that system \cite{Harris2024}. Likewise, instead of using pressure, substrate temperature, and fluence to span the growth parameter space, substrate temperature and the respective temperatures of each effusion cell may sufficiently describe the growth conditions of MBE growth, for example. Depending on the use case, future applications of the methodology could utilize more adaptive surrogate models to capture phase changes. Or they could utilize more exploratory acquisition functions to obtain a more detailed phase mapping of the growth parameter space. Furthermore, instead of summing weighted individual metrics, using other functional forms such as the logarithm or higher degree polynomials may better predict and quantify film quality. For example, if one measured characteristic — \textit{e.g.}, roughness —  exhibits large outliers, using the logarithm of that variable will dampen the effect of outliers on the GP surrogate model while increasing model sensitivity to small variations of measured values. As experimental automation advances, multi-modal film characterization combined with machine learning-driven synthesis portends new opportunities for synthesizing complex oxide films of higher quality and reproducibility.

\section{Methods}

\subsection{Target synthesis, substrate preparation and film growth} A LVO target was prepared by pressing and firing stoichiometric amounts of La and V at 1000 $\degree$C for 24 hr in 4\% H$_2$ + Ar to reduce the target from LaVO$_4$ to LaVO$_3$. The substrates used were (001) STO which were 5 × 5 mm$^2$, single-side polished, miscut 0.025–0.050$\degree$, CrysTec GmbH and were adhered to the heating stage wth Ag paint. Prior to mounting, the substrates were etched in 10:1 HF for 1 min, rinsed in DI water, dried under N$_2$, and then annealed at 1100 $\degree$C for 1 h (300 $\degree$C h$^{-1}$ ramp) to obtain nominal TiO$_2$ termination. Absorption measurements required double-side-polished (DSP) STO substrates, which were mechanically clamped to the heating stage rather than adhered with Ag paint. Reduced contact with the plate decreased the effective thermal conductivity, which in turn decreased the available temperature range to $<$700 $\degree$C.  LVO films were grown by PLD using a KrF excimer laser (248 nm, 25 ns, 5 Hz; Coherent LPX 305) to ablate the LVO target. A projection optical beam path imaged a rectangular aperture onto the target, yielding a 1.32 × 1.36 mm$^2$ spot (0.0180 cm$^2$). Each film was deposited with 7500 laser pulses at a 50 mm target-to-substrate distance. 

The range of fluence, temperature, and PO$_2$ studied for growth were 0.8 - 2.2 J/cm$^2$, 500 - 830 $\degree$C, and 3$\times$10$^{-8}$ - 2$\times$10$^{-2}$ Torr, respectively. Laser pulse energy was measured inside the chamber before each growth. To measure temperature during growth, we used a pyrometer directed at the edge of the substrate and calculated the average temperature throughout growth. To calculate PO$_2$ values, we used a residual gas analyzer and integrated pressure values over 31.5-32.5 amu for total pressure values $<$ 5$\times$10$^{-5}$ Torr and subtracted working pressure values from base pressure values for total pressure values $>$ 5$\times$10$^{-5}$ Torr. To reach PO$_2$ values $\geq$ 4$\times$10$^{-7}$ Torr, we varied flow rates of oxygen gas until the pressure stabilized. All growths began with a vacuum environment on the order of 10$^{-6}$ Torr.

\subsection{Structural Characterization} XRD was conducted with a PANalytical X’Pert Pro MRD diffractometer for $2\theta-\omega$, rocking curve, and reciprocal space mapping measurements. $2\theta-\omega$ scans range from 20 - 80$\degree$ and 43 - 51$\degree$, using a 0.02 $\degree$/step with 1 s integration time and 0.005$\degree$/step and 2 s integration time, respectively. To determine the out-of-plane lattice parameter of films, we used data from the 20 - 80$\degree$ scans and fit the (002) film and substrate peaks using a film-substrate model which accounted for their respective $c$ values and thickness fringes \cite{10.1063/1.120377, I_K_Robinson_1992}. To determine the $\Delta\omega$ of film peaks, we acquired rocking curves with 0.01$\degree$ step size and 1 s integration time around the (002) peak positions of films and fit the data to a pseudo-Voigt function. We obtained values for integrated LaVO$_4$ intensity, $I_{LaVO_4}$, from the narrow 43 - 51$\degree$ scans, normalizing data to the maximum substrate intensity, multiplying by 10$^6$ to obtain values near unity, and using the \texttt{numpy.trapz} function from NumPy to integrate over 48 - 51$\degree$ \cite{harris2020array}. LaVO$_4$ is structurally monoclinic with a wide-band gap, and the peak at 49.5$\degree$ corresponds to its (040) reflection \cite{LaVO4_gap, Hotta2006}. LaVO$_4$ forms at the surface of LVO at ambient conditions and grows instead of LVO at high PO$_2$. High-temperature XRD (HTXRD) measurements were done with a PANalytical X’Pert Pro MPD XRD system by increasing substrate temperature up to 800 $\degree$C, flowing 50-70 sccm of 4\% H$_2$ + Ar, and using 0.02$\degree$/ step and 2 s integration time. We measured the surface roughness with a Bruker Dimension Icon AFM in tapping mode with a Si probe (TESPA-V2, 7 nm tip radius, 37 N m$^{-1}$ spring constant). We acquired the $R_{RMS}$ of 3$\times$3 $\mu$m scans operated at 2 Hz and 512 samples/line.

\subsection{GPBO Setup} Variational Gaussian process regression was implemented via GPax \cite{ziatdinov2021hypothesis, ziatdinov2021physics} using a Mat{\'e}rn kernel with a zero-mean function and was trained via stochastic variational inference with the evidence lower bound loss criterion. We placed LogNormal(0,1) priors on the noise variance and lengthscales. For more information on the implementation of the setup above, \autoref{math_GPBO} in the supplement provides a discussion. Based on experimental constraints and the reported values for optimum growth conditions in the literature [\autoref{table:PLD_param_table}], the bounds of the growth parameter space in the GPBO were set to fluence = 0.8 - 2.2 J/cm$^2$, temperature = 500 - 835 $\degree$C, and PO$_2$ = 3$\times$10$^{-8}$ - 3$\times$10$^{-4}$ Torr. The growth parameter space for the GPBO algorithm was discretized in fluence, temperature and PO$_2$ from 0.8 - 2.2 J/cm$^2$ with 0.1 J/cm$^2$ steps, 500 - 835 $\degree$C with 20$\degree$C steps, and 3$\times$10$^{-8}$ - 3$\times$10$^{-4}$ Torr or -7.5 - -3.5 with 0.1 steps on a log$_{10}$ scale. The uncertainty of the growth parameter values along with experimental intuition dictated the step sizes. For example, a fluence of 1.0 J/cm$^2$ translated to pulse energies of 18.0 mJ, and the uncertainty of pulse energies was $\pm$ 0.5 mJ. Therefore, we could consistently differentiate increments of 0.1 J/cm$^2$. Similarly, temperature steps of 20 $\degree$C were used because close values would not yield meaningful differences (\textit{e.g.} 603 $\degree$C vs. 600 $\degree$C). We did not include samples grown at PO$_2$ $>$10$^{-3}$ Torr in the GPBO algorithm because they were pure polycrystalline LaVO$_4$ with no detectable LVO phase. We used the expected improvement acquisition function to balance exploration and exploitation until the global minimum stabilized at some value and the maximum of the expected improvement normalized to the GP mean reached $\sim$0.05.

\subsection{Optical Characterization} For a subset of samples grown near and away from the low-temperature growth mode identified by  the GPBO approach, optical properties were measured to elucidate relationships between functional properties and indicators or film quality. The absorption measurements involved DSP (001) STO from MTI corporation. Each sample received a different number of shots such that film thicknesses were approximately 100 nm. Measurements were done in transmission using a Cary 5000 UV-Vis-NIR spectrophotometer (Agilent Technologies) scanning from 200-1500 nm with a scan rate of 600 nm/s and 12 nm/s below and above 800 nm, respectively. Fits to the absorption data were acquired by summing four Gaussians centered at electronic transitions found in the literature and by adding a constant term to account for sub band gap absorption due to defects \cite{Wang2015}. Tauc plot fitting required multiplying the absorption coefficent by the incident light energy and squaring (square-rooting) this quantity for direct (indirect) transitions. Using this quantity, the direct (indirect) gaps were acquired by linear fitting about the transition peaks, found through the absorption spectra.

\section{Acknowledgment}
This work was supported by the U. S. Department of Energy (DOE), Office of Science, Basic Energy Sciences (BES), Materials Sciences and Engineering Division (machine learning analysis, structural and optical characterization and analysis). The algorithm development was supported by the Center for Nanophase Materials Sciences (CNMS), which is a U.S. Department of Energy, Office of Science User Facility, at Oak Ridge National Laboratory. This work was also supported by the Laboratory Directed Research and Development Program (LDRD) of Oak Ridge National Laboratory, managed by UT-Battelle, LLC, for the U.S. Department of Energy (some analysis). This material is based upon work supported by the DOE, Office of Science, Office of Workforce Development for Teachers and Scientists, Office of Science Graduate Student Research (SCGSR) program.

\section{Data Availability}

Data sets generated during the current study are available from the corresponding author on reasonable request.

\section{Supplemental Information}

The Supplement includes PLD parameters from the literature (\autoref{table:PLD_param_table}), more information on the mathematical construction of the GPBO algorithm (\autoref{math_GPBO}), plots of linear correlation between individual measurements and the distribution of measured values (\autoref{fig:supp_corre}), all plots of slices of the GP surrogate model mean and its gradient (\autoref{supp_gp_grad} and \autoref{fig:supp_full_sens_plots}), $2\theta$-$\omega$ XRD data as function of annealing temperature (\autoref{fig:htxrd_many_scans} and \autoref{fig:htxrd_before_after_no_Pt}), an estimation of the out-of-plane lattice parameter as a function of Poisson ratio (\autoref{fig:supp_poisson}), and Tauc plots for samples not pictured in the main text (\autoref{tauc_b41} and \autoref{tauc_b42}).

\printbibliography

\pagebreak

\clearpage
\setcounter{figure}{0}
\renewcommand{\thefigure}{S\arabic{figure}}

\setcounter{section}{0}
\setcounter{subsection}{0}
\setcounter{secnumdepth}{2}   

\renewcommand{\thesection}{S\arabic{section}}
\renewcommand{\thesubsection}{S\arabic{section}.\arabic{subsection}}

\setcounter{table}{0}
\renewcommand{\thetable}{S\arabic{table}}

\section*{Supplement}

\section{PLD Parameters from Literature}

\begin{table}[h]
\resizebox{\textwidth}{!}{%
    \begin{tabular}{|c|c|c|c|c|c|c|c|c|}
    \hline
        Ref. &Target & Temp. ($^{\circ}$C) &	PO$_2$ (Torr)	& Fluence (J/cm$^{2}$) &	Rep. Rate (Hz) & Pressure (Torr) \\

\hline
        \normcite{Hotta2006} &LaVO$_4$ & 500-900 &	10$^{-8}$ - 10$^{-2}$	& 2.5 &	4 & 10$^{-8}$ - 10$^{-2}$ \\

\hline
        \normcite{Wang2015} &LaVO$_4$ & 700 &	$<$10$^{-6}$	& 2 & 10 & 10$^{-6}$  \\

\hline
       \normcite{Wadehra2020} &LaVO$_4$ & 600 &	10$^{-6}$	& 4 &	2 & N/A \\

\hline
        \normcite{Zhang2021} &LaV$_{1.2}$O$_4$ & 600-650 &	N/A & 0.6 - 2.0 & 1 & 10$^{-10}$ \\

\hline
        \normcite{Masuno2004} &LaVO$_3$ & 700 &	N/A & 1 & 3-5 & 4$\times$10$^{-6}$\\

\hline
        \normcite{Rotella2012} &LaVO$_4$ & 700 &	N/A & 2 & 3 & 7.5$\times$10$^{-6}$\\

\hline
        \normcite{Cheikh2024} &LaVO$_4$ & 600-800 &	7.5$\times$10$^{-7}$ - 7.5$\times$10$^{-5}$ & N/A & 2 & N/A\\

\hline
        \normcite{Meley2018} &LaVO$_4$ & 800-900 &	3.75$\times$10$^{-7}$ & 2 & 1 & N/A\\

\hline
        \normcite{Higuchi2009} &LaVO$_4$ & 600 &	1$\times$10$^{-6}$ & 1 & 8 & N/A\\

    \hline
    \end{tabular}}
    \caption{PLD Growth Parameters from the literature. For optimal growth, \normcite{Hotta2006} reported 600$^{\circ}$C and 10$^{-6}$ Torr. \normcite{Cheikh2024} grew 2 series of samples, one varying temperature and the other PO$_2$.}
    \label{table:PLD_param_table}
\end{table}

\section{Mathematical Construction of the GPBO Algorithm}
\label{math_GPBO}

To understand the active learning algorithm mathematically, consider the latent function $f$, which is unknown but perfectly represents the dependence of the parameters (|$c -3.945$|, $\Delta\omega$, $I_{LaVO_4}$, and $R_{RMS}$) on the growth variables (fluence, pressure and temperature). The goal is to determine and approximate the form of $f$ given noisy measurements $y_i$ in the domain of the growth parameter space $(\boldsymbol{x})$. To do so, a GP prior $f \sim \mathcal{GP}[m(\cdot), k_\theta(\cdot,\cdot)]$ with zero mean ($m$) and a Matérn kernel ($k_\theta$) is used. That is, it is assumed a Gaussian process accurately describes $f$, and the Matérn kernel given some hyperparameters $\theta$ describes the covariance of $f$ for two points in the growth parameter space: $\boldsymbol{x}$ and $\boldsymbol{x'}$. Furthermore, given a Gaussian noise component ($\epsilon_i$) with zero mean and homoscedastic variance $\sigma_n^2$ for each observation, each sample measurement takes the form:

\begin{equation}
y_i = f(\boldsymbol{x}_i) + \epsilon_i,\quad \epsilon_i \sim \mathcal{N}(0,\sigma_n^2).
\label{eq:obs_model}
\end{equation}

Given data $\mathcal{D}=\{(\boldsymbol{x}_i,y_i)\}_{i=1}^N$ and using Bayes theorem, the posterior distribution function at a new $\boldsymbol{x}$ is Gaussian,
\begin{equation}
p\!\left(f(\boldsymbol{x}) \mid \mathcal{D}\right)=\mathcal{N}\!\big(\mu(\boldsymbol{x}),\,\sigma^2(\boldsymbol{x})\big),
\label{eq:posterior}
\end{equation}
with $\mu(\boldsymbol{x})=\mathbb{E}[f(\boldsymbol{x})\mid\mathcal{D}]$ and $\sigma^2(\boldsymbol{x})=\mathrm{Var}[f(\boldsymbol{x})\mid\mathcal{D}]$ available in closed form. Physically, $\mu(\boldsymbol{x})$ models the weighted sum of |$c$ - 3.945|, $\Delta\omega$, $I_{LaVO_4}$, and $R_{RMS}$ throughout the growth parameter space, and $\sigma^2(\boldsymbol{x})$ gives the uncertainty of $\mu(\boldsymbol{x})$.

For Bayesian optimization the Expected Improvement (EI) algorithm is used. For minimization, let $y^{*}=\min_i y_i$ be the best observed value and define the improvement
\begin{equation}
I(\boldsymbol{x})=\max\!\big(y^{*}-f(\boldsymbol{x}),\,0\big).
\label{eq:improvement}
\end{equation}
With
\begin{equation}
z(\boldsymbol{x})=\frac{y^{*}-\mu(\boldsymbol{x})}{\sigma(\boldsymbol{x})}
\label{eq:z}
\end{equation}
and $\Phi$ and $\phi$ the standard normal CDF and PDF, respectively, the EI has the closed form
\begin{equation}
\mathrm{EI}(\boldsymbol{x})=
\begin{cases}
\big(y^{*}-\mu(\boldsymbol{x})\big)\,\Phi\!\big(z(\boldsymbol{x})\big)+\sigma(\boldsymbol{x})\,\phi\!\big(z(\boldsymbol{x})\big), & \sigma(\boldsymbol{x})>0,\\[0.2em]
0, & \sigma(\boldsymbol{x})=0.
\end{cases}
\label{eq:ei}
\end{equation}
The next experiment is chosen by maximizing $\mathrm{EI}(\boldsymbol{x})$ (see \normcite{alma991044472453103276} for more details). In summary, after assuming the form of $f$ and $\epsilon_i$ and finding the posterior distribution function $p\!\left(f(\boldsymbol{x}) \mid \mathcal{D}\right)$, $\mu(\boldsymbol{x})$ interpolates the weighted sum of individual measurements and $\sigma^2(\boldsymbol{x})$ quantifies the uncertainty of the interpolation throughout the growth parameter space ($\boldsymbol{x}$).











\section{Correlations of Individual Metrics}

\begin{figure}[H]
  \centering
  \includegraphics[width=\linewidth,
                   trim=0cm 0cm 0cm 0cm, clip]{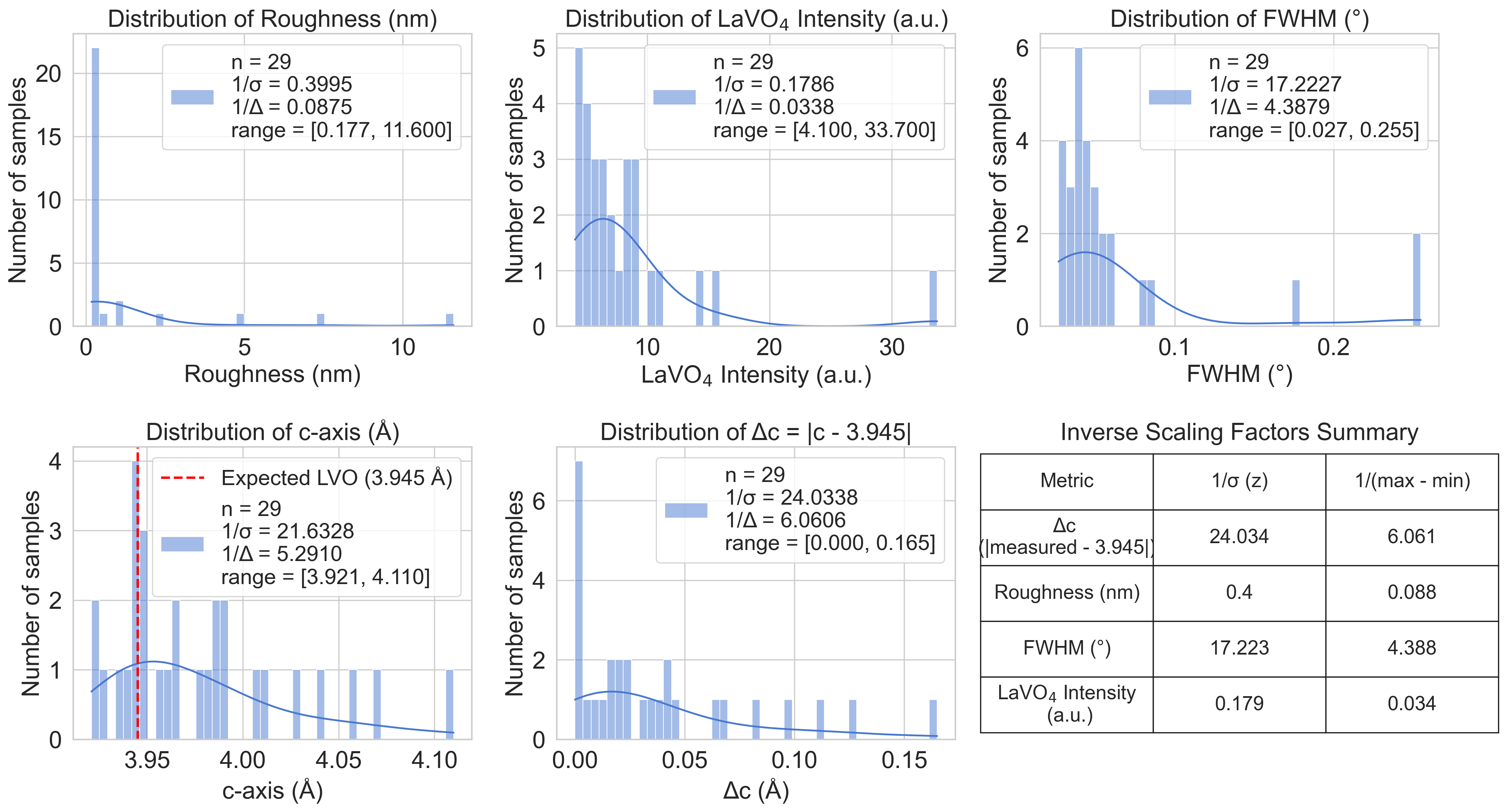}
  \caption{Distributions of scores from individual characterization measurements. 29 samples were included in the final GP surrogate model. The table gives z-score and min-max scaling values from the 29 samples. Actual min-max scaling values used for GP model are slightly different because they were obtained for an earlier iteration of sample growth and characterization. Obtaining scaling values from characterization experiments are necessary right after inputting seed points to obtain reasonable predictions for future sample growths that properly integrate each type of measurement.}
  \label{fig:Supp_distr}
\end{figure}

\begin{figure}[H]
    \centering
    \begin{subfigure}[b]{\linewidth}
        \centering
        \includegraphics[width=\linewidth]{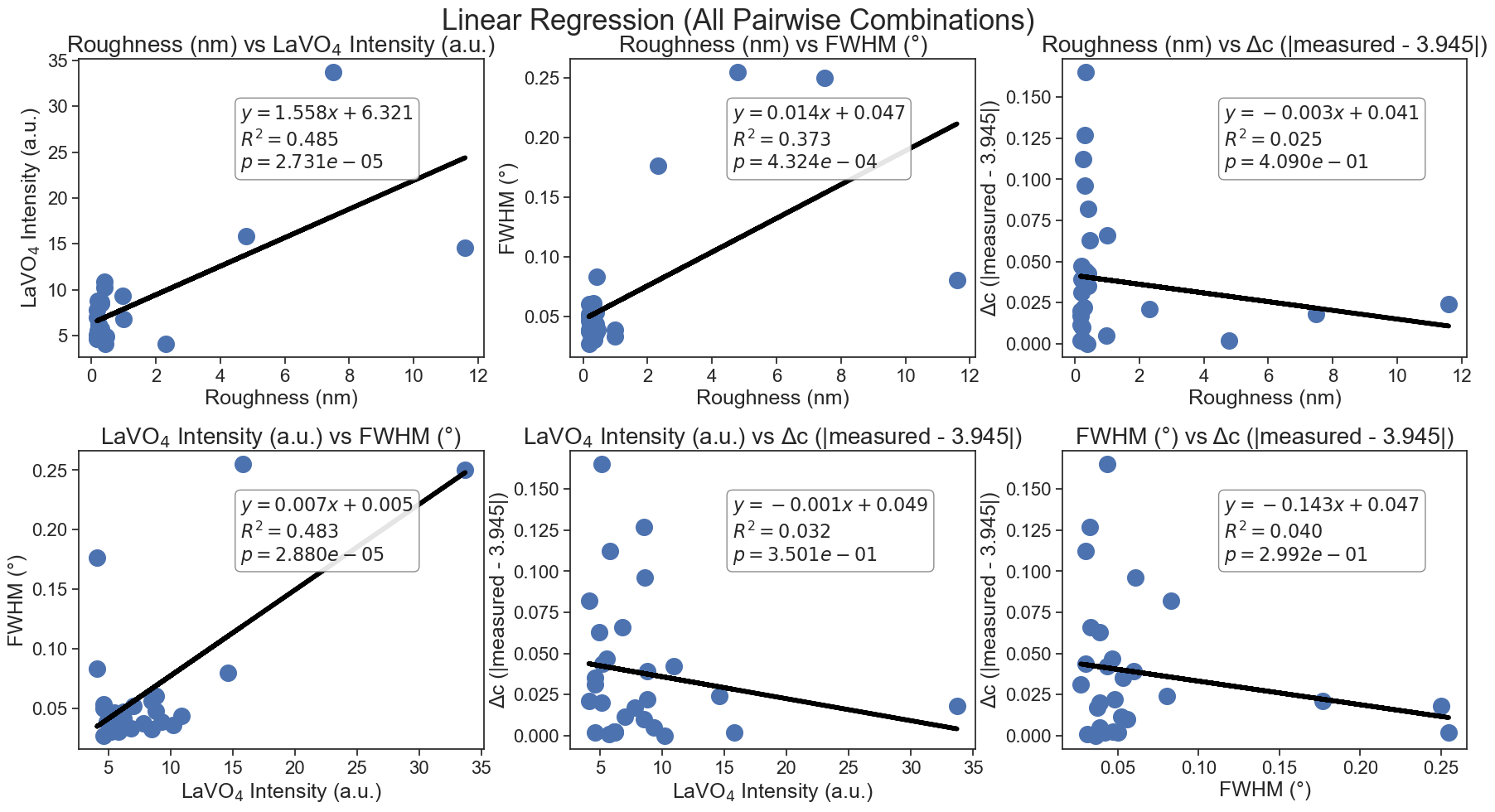}
        \caption{Linear correlations.} 
        \label{fig:linear_corr}
    \end{subfigure}

    \vspace{1em} 


    \caption{Pairwise correlations of values from individual characterization measurements. Deviation from target out-of-plane lattice parameter is weakly anti-correlated to other metrics. The other 3 measurements are moderately correlated.}
    \label{fig:supp_corre}
\end{figure}

\section{All Plots of Slices of GP Surrogate Mean and its Gradient}
\label{supp_gp_grad}

\begin{figure}[H]
    \centering
    \includegraphics[width=\linewidth,trim=0cm 0cm 0cm 0.5cm, clip]{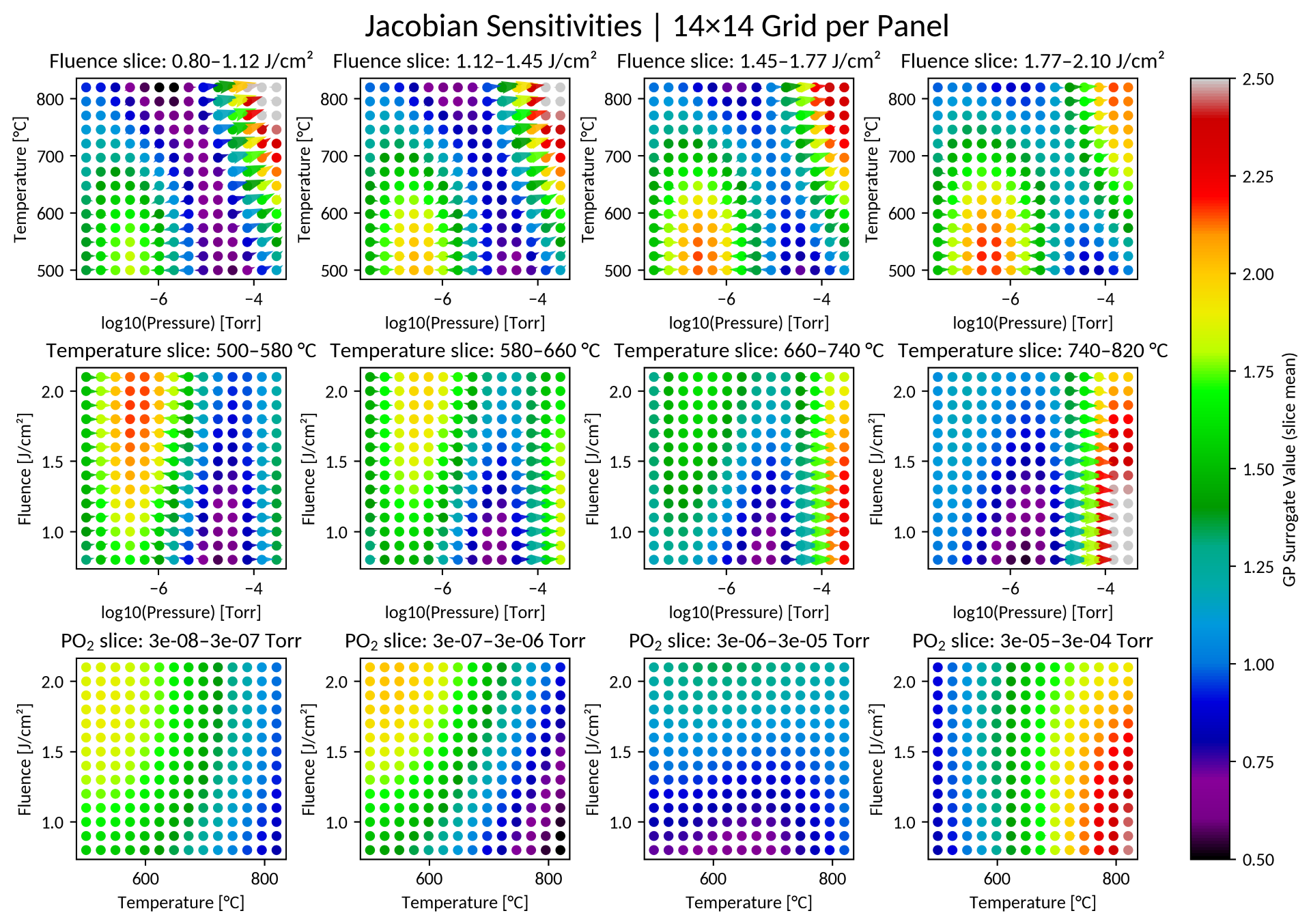}
    \caption{The plots show slices of the full 3D surrogate model. The dot colors are the average values of the GP surrogate model mean for the respective slices. Since the plots visualize the GP mean with 14 points along each dimension and there are 4 slices per dimension, the GP Mean value at a point in some slice is the average GP mean over 3-4 points. A 14x14 grid is only chosen because it was found to be the best way to visualize the data. The arrows correspond to the Jacobian of the GP mean, or the gradient in this case since the GP surrogate model mean is a scalar-valued function. The arrows point in the direction of the gradient. The arrow sizes correspond to the magnitude of the gradient at each point and are also the average gradient values for the slice. The largest arrows appear at low fluence, high temperature and high PO$_2$. In this region, the sample growth landscape quickly transitions from the global optimum to mixed-phase films that received poor sample scores. Arrows are smaller than the dot size for some points, especially when PO$_2$ is held constant across a slice, indicating little influence on the GP-mean surface.}
    \label{fig:supp_full_sens_plots}
\end{figure}

This data-driven, active-learning approach facilitates conclusions about film growth regions that would have been difficult to find intuitively and elucidates complex mechanisms of epitaxial film growth by PLD. PLD involves a complicated interplay of growth parameters, especially pressure and fluence since plume dynamics depend on both. To study the impact of PO$_2$, temperature, and fluence on the mean of the GP surrogate model, we find the gradient of the GP mean. This calculation is shown in \autoref{fig:supp_full_sens_plots}, where the arrow colors correspond to values of the GP mean after the last growth iteration. The global optimum (820 $\degree$C, 0.8 J/cm$^2$, and 1.2$\times$10$^{-6}$ Torr) is easily visualized at the slice of fluence = 0.80 - 1.15 J/cm$^2$ and is relatively stable, indicated by vector values with small magnitudes. In the regime of high temperature, high PO$_2$, and low fluence, this technique captures the large dependence of the GP mean on PO$_2$, indicating that it drives the transition to the mixed LVO-LaVO$_4$ phase. In fact, the GP surrogate model is roughly twice as sensitive to log$_{10}$ (PO$_2$) than temperature throughout the growth parameter space, which is indicated by values of the gradient projected onto the temperature, PO$_2$, or fluence axis. \autoref{fig:sens_table} outlines this behavior, giving gradient projections for different fluence, temperature and PO$_2$ slices. Plots of the gradient for all the different fluence, temperature, and PO$_2$ slices, shown in \autoref{fig:supp_full_sens_plots}, also reflect the importance of PO$_2$ in driving the film quality landscape and provide insights into synthesis mechanisms of other growth regimes. 

Throughout the GP mean model, fluence marginally affects the film quality landscape except for moderate PO$_2$ and low temperature, as seen in \autoref{fig:sens_table}. This table shows the average value of the gradient projected onto the PO$_2$, temperature, or fluence axis for slices of the GP surrogate model. In the regime of moderate PO$_2$ and low temperature, fluence affects the ratios of cationic species to oxygen without the dominating effects on film quality of high PO$_2$, which drives the LaVO$_4$ phase, or high temperature, which likely heals defects and could introduce oxygen to the film through the STO. At low temperatures, the incident kinetic energy of the growth species, which the fluence determines, constitutes a greater share of the internal energy and lateral momentum of the growing film. Therefore, fluence drives the arrangement of adatoms into a crystalline structure in this region more than other ones. Overall, the choice of a logarithmic scale for PO$_2$ likely causes the marginal effect of fluence on the GP model since pressure exponentially attenuates the propagation of plume species. At a typical PLD total pressure (10$^{-3}$-10$^{-1}$ Torr), the dependence of film growth on fluence would likely be higher since the plume shape, mean-free path of species, and relative velocities of fast and slow plume components affect growth dynamics \cite{Geohegan1996}.

\begin{table}
    \centering
    \includegraphics[width=\linewidth]{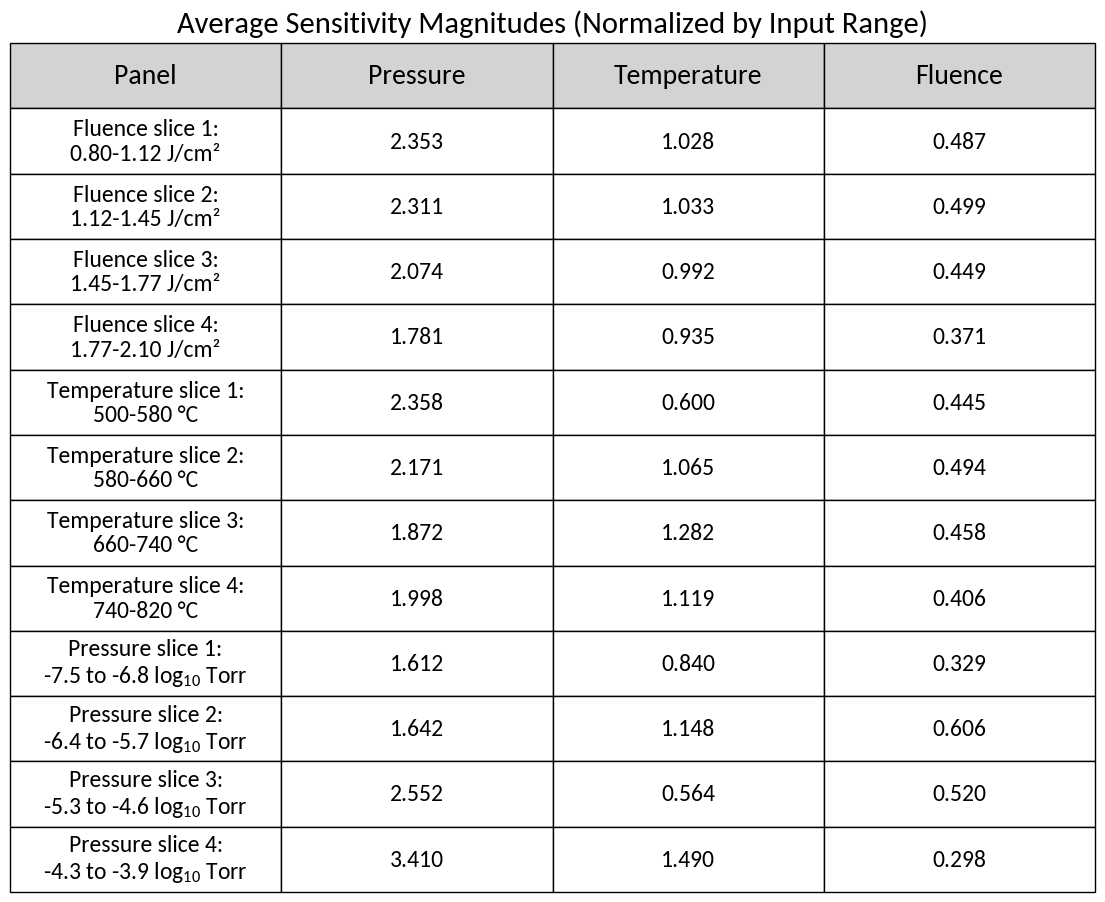}
    \caption{Table of average gradient projections for each slice normalized by the length of respective axes (axis$_{max}$ - axis$_{min}$). We compute the gradient of the GP surrogate model, the average value of the gradient projected onto each axis for each slice, and then normalize those values by the length of respective axes. Although all we compute here is the gradient of the GP surrogate model mean, our code generalizes to vector-valued GP surrogate models by calculating the Jacobian of the model [GitHub repository: \url{https://github.com/J-dot-hue/Optimizing-Pulsed-Laser-Deposition-of-LaVO3-by-Active-Learning}]. In other words, the approach is suited to multivariate Gaussian surrogate models also known as multi-task learning. For example, an experimenter may want to explicitly separate film characterization metrics for morphology, stoichiometry, and optical properties into a three-component surrogate model.}
    \label{fig:sens_table}
\end{table}

\section{HTXRD Data}

\begin{figure}[H]
    \centering
    \includegraphics[width=\linewidth]{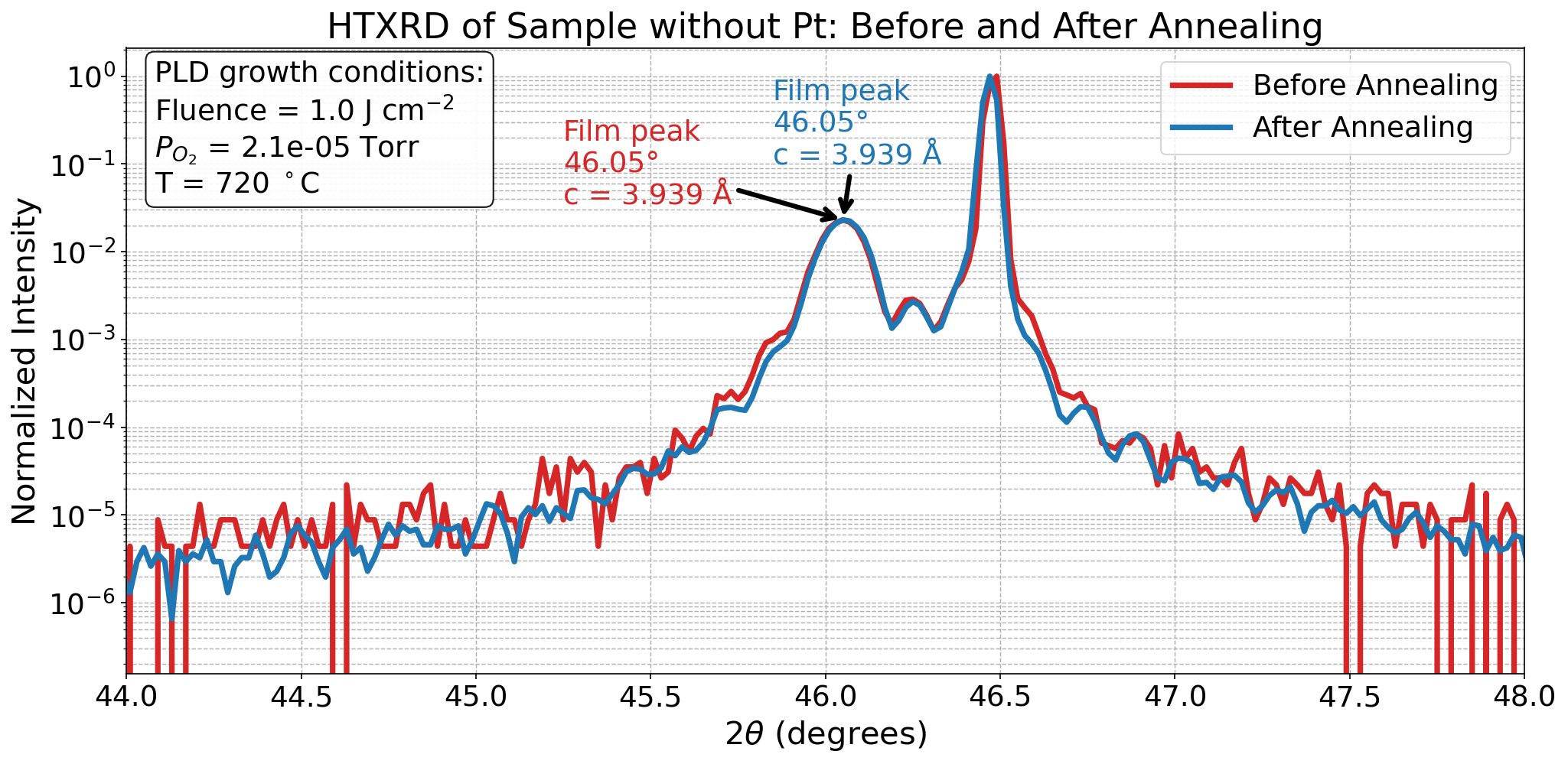}
    \caption{XRD data before and after annealing a sample up to 800 $^{\circ}$C in a reducing environment. The data show virtually no change in $c$.}
    \label{fig:htxrd_before_after_no_Pt}
    \includegraphics[width=\linewidth]{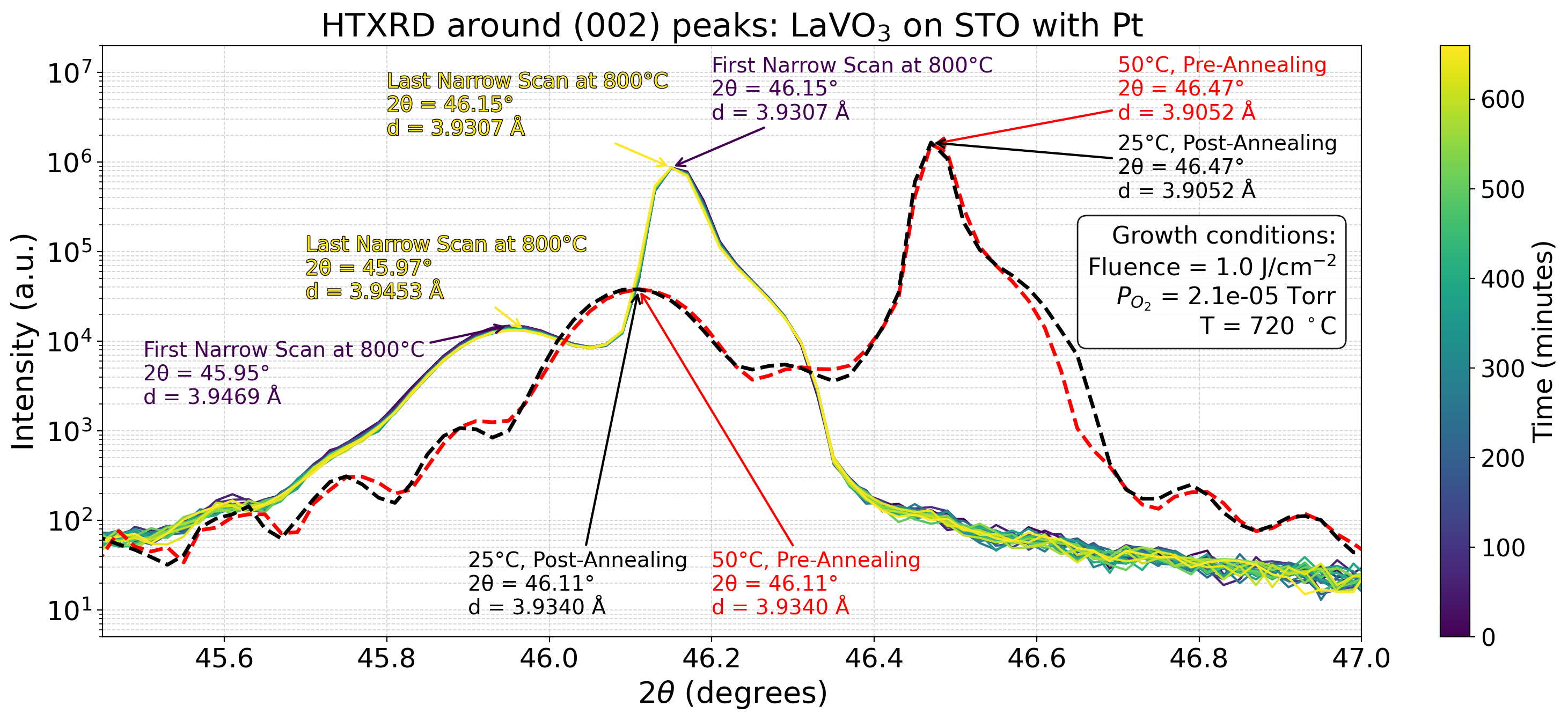}
    \caption{After depositing Pt on the sample and executing a set of anneals up to 800 $^{\circ}$C, XRD measurements were taken before, after, and during annealing in a reducing environment at 800 $^{\circ}$C. Before and after annealing, the $c$ stays virtually unchanged. During annealing the LVO and STO $c$ change due to thermal expansion. The LVO out-of-plane lattice parameter stays virtually unchanged during the anneal.}
    \label{fig:htxrd_many_scans}
\end{figure}

\section{Out-of-Plane Lattice Parameter and Poisson Ratio}
    \label{poisson_c}

 LVO has a GdFeO$_3$ structure with room-temperature lattice constants: a $\approx$ 5.555 Å, b $\approx$ 7.849 Å, and c $\approx$ 5.553 Å \cite{Martnez-Lope2008, Khan2004, Nguyen1995, Miyasaka2002, Bordet1993, Onoda1996}. Therefore, the half-unit cell, pseudocubic lattice parameter of bulk LVO is approximately 3.925 Å \cite{Rotella2015StructuralFilms}.

 \begin{figure}[H]
     \centering
     \includegraphics[width=\linewidth,
                   trim=0cm 0cm 0cm 0cm, clip]{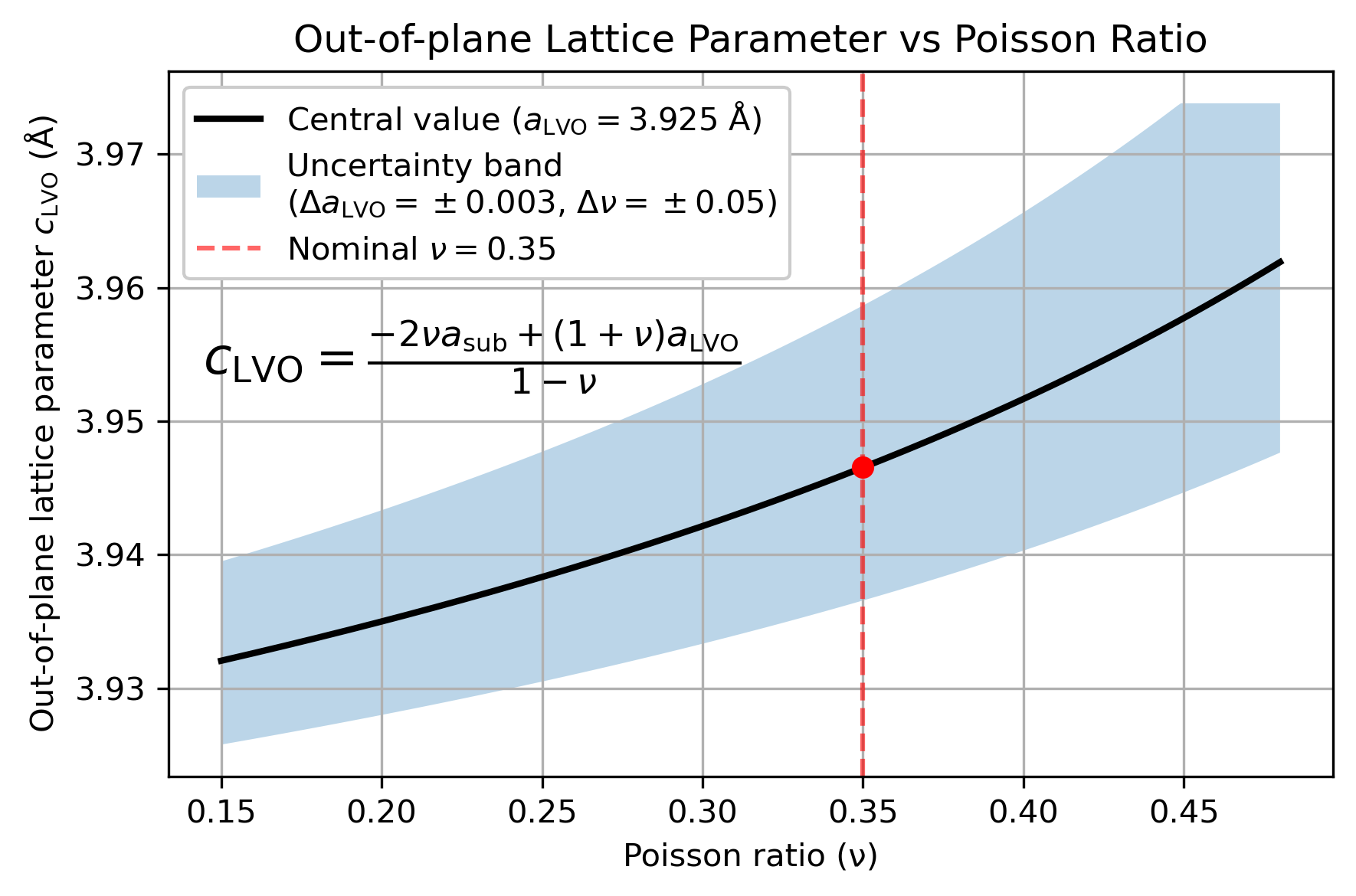}
     \caption{Out-of-plane lattice parameter as a function of Poisson ratio, which accounts for epitaxial strain. The maximum $c_{LVO}$ of the uncertainty band is obtained by plugging in nominal values for $a_{LVO}$ and $\nu$ plus their respective uncertainty values. The minimum $c_{LVO}$ of the uncertainty band is obtained by plugging in nominal values for $a_{LVO}$ and $\nu$ minus their respective uncertainty values.}
     \label{fig:supp_poisson}
 \end{figure}

 \section{Tauc Plots for other two samples not included in main text}

The first three plots correspond to the purple curve in \autoref{fig:Figure_5_labels}, the sample with a large $c$ but strong scores otherwise. The bottom three plots correspond to the green curve, the sample near the target $c$ but poor scores otherwise. The y-axis units are in (eV $\cdot$ nm$^-1$)$^2$ for the direct gap plots and in units of (eV $\cdot$ nm$^-1$)$^{1/2}$ for the indirect gap plots.

\begin{figure}
    \begin{overpic}[width=\linewidth,
                   trim=0cm 0cm 17.5cm 1cm, clip]{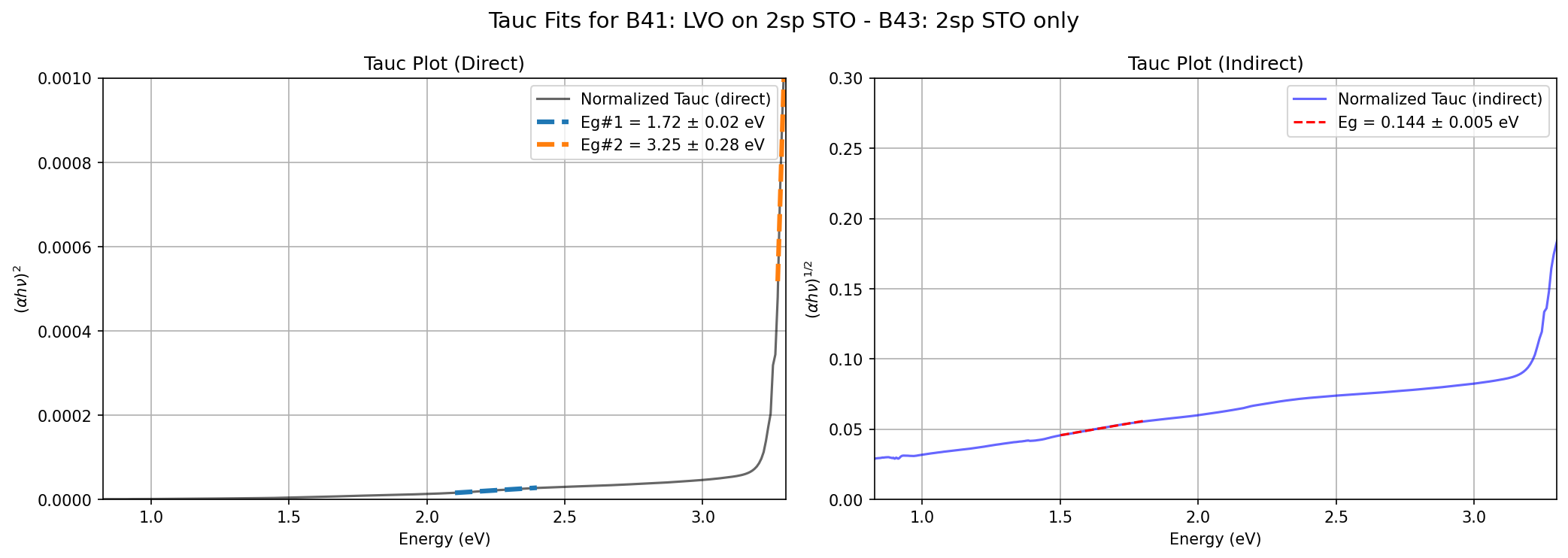}
        \put(20,40){%
        \parbox{4cm}{\large f = 1.5 J/cm$^2$\\
                            PO$_2$ = 3.6$\times$10$^{-6}$\\
                            T = 655 $\degree$C}%
}
    \end{overpic}

    \begin{overpic}[width=\linewidth,
                   trim=18cm 0cm 0cm 1cm, clip]{figures/b41-b43_tauc_1.png}
        \put(20,40){%
        \parbox{4cm}{\large f = 1.5 J/cm$^2$\\
                            PO$_2$ = 3.6$\times$10$^{-6}$\\
                            T = 655 $\degree$C}%
}
    \end{overpic}
\end{figure}

\begin{figure}
    \begin{overpic}[width=\linewidth,
                   trim=0cm 0cm 17.5cm 1cm, clip]{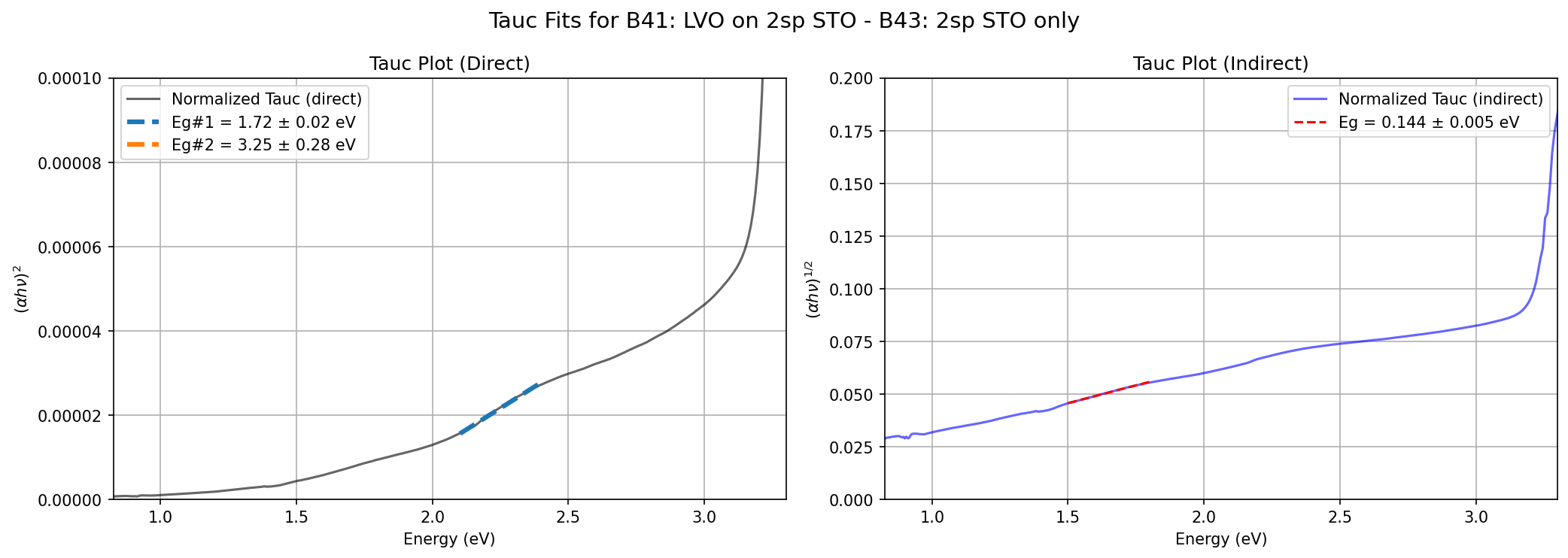}
        \put(20,40){%
        \parbox{4cm}{\large f = 1.5 J/cm$^2$\\
                            PO$_2$ = 3.6$\times$10$^{-6}$\\
                            T = 655 $\degree$C}%
}
    \end{overpic}
    \caption{All three plots correspond to the sample data given by the purple curve in \autoref{fig:Figure_5_labels}. Two direct gap Tauc plots are given. One is zoomed in on the data used to acquire one of the fits for the direct gap.}
    \label{tauc_b41}
\end{figure}

\begin{figure}
    \begin{overpic}[width=\linewidth,
                   trim=0cm 0cm 17.5cm 1cm, clip]{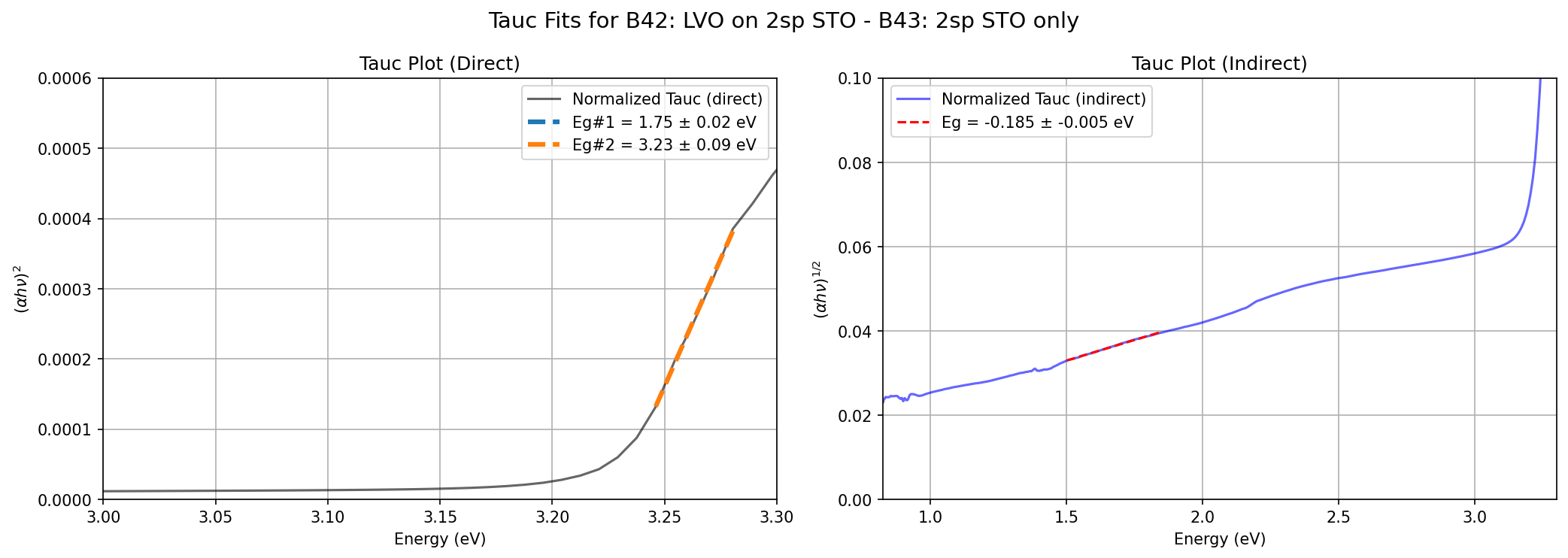}
        \put(20,40){%
        \parbox{4cm}{\large f = 1.3 J/cm$^2$\\
                            PO$_2$ = 1.4$\times$10$^{-4}$\\
                            T = 635 $\degree$C}%
}

    \end{overpic}

    \begin{overpic}[width=\linewidth,
                   trim=18cm 0cm 0cm 1cm, clip]{figures/b42-b43_tauc_1.png}
        \put(20,40){%
        \parbox{4cm}{\large f = 1.3 J/cm$^2$\\
                            PO$_2$ = 1.4$\times$10$^{-4}$\\
                            T = 635 $\degree$C}%
}

    \end{overpic}

\end{figure}

\begin{figure}
    \begin{overpic}[width=\linewidth,
                   trim=0cm 0cm 17.5cm 1cm, clip]{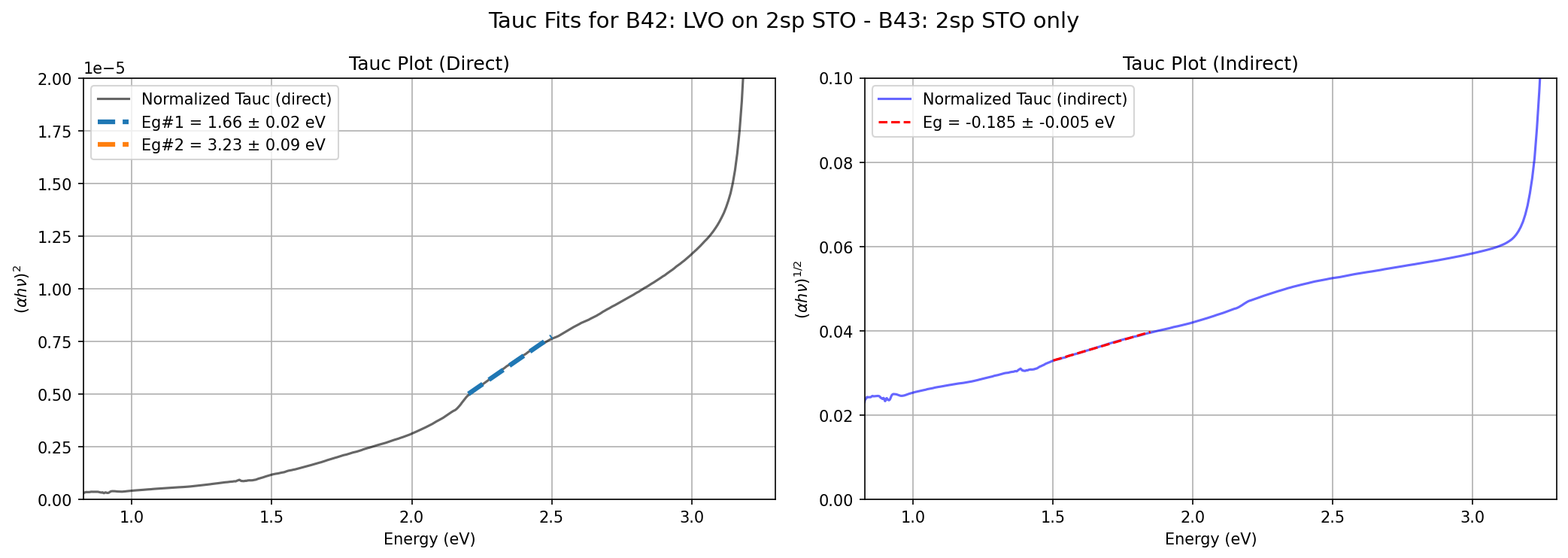}
        \put(20,40){%
        \parbox{4cm}{\large f = 1.3 J/cm$^2$\\
                            PO$_2$ = 1.4$\times$10$^{-4}$\\
                            T = 635 $\degree$C}%
}

    \end{overpic}
    \caption{All three plots correspond to the sample data given by the green curve in \autoref{fig:Figure_5_labels}.  Two direct gap and one indirect gap Tauc plots are given. Both are zoomed in on the data used to acquire one of the respective fits for the direct gap.}
    \label{tauc_b42}
\end{figure}

\end{document}